\documentclass[prd,twocolumn,nofootinbib]{revtex4-1}
\usepackage{graphicx}
\usepackage{bm}
\usepackage[colorlinks=true]{hyperref}
\usepackage{float}
\usepackage{amsmath}
\usepackage{amssymb}
\usepackage{pifont}
\usepackage{gensymb}
\usepackage{multirow}
\usepackage[dvipsnames]{xcolor}
\usepackage[utf8]{inputenc}
\usepackage{subcaption}

\newcommand\be{\begin{equation}}
\newcommand\ba{\begin{eqnarray}}
\newcommand\ee{\end{equation}}
\newcommand\ea{\end{eqnarray}}
\newcommand\bw{\begin{widetext}}
\newcommand\ew{\end{widetext}}

\newcommand{\KN}{{\mbox{\tiny KN}}}
\newcommand{\ISCO}{{\mbox{\tiny ISCO}}}

\newcommand{\PPN}{{\mbox{\tiny PPN}}}

\definecolor{KellyGreen}{RGB}{76,187,23}

\begin{document}
\title{Rotating black holes in valid vector-tensor theories after GW170817}
\author{Siddarth Ajith}
\affiliation{Department of Physics, University of Virginia, Charlottesville, Virginia 22904, USA}
\author{Alexander Saffer}
\affiliation{Department of Physics, University of Virginia, Charlottesville, Virginia 22904, USA}
\author{Kent Yagi}
\affiliation{Department of Physics, University of Virginia, Charlottesville, Virginia 22904, USA}

\date{\today}
\begin{abstract}
  Vector-tensor theories beyond General Relativity have widely been studied in the context of ultraviolet completion of gravity, endowing a mass to the graviton and explaining dark energy phenomena. 
  We here construct rotating black hole solutions in vector-tensor theories valid after the binary neutron star merger event GW170817 that placed very stringent bound on the propagation speed of gravitational waves away from the speed of light. 
  Such valid vector-tensor theories are constructed by performing a generic conformal transformation to Einstein-Maxwell theory, and the new rotating black hole solutions are constructed by applying the same conformal transformation to the Kerr-Newman solution.     
  These theories fall outside of beyond generalized Proca theories but are within an extended class of vector-tensor theories that satisfy a degenerate condition to eliminate instability modes and are thus healthy. 
  We find that such conformal Kerr-Newman solutions preserve the location of the singularities, event horizons and ergoregion boundary from Kerr-Newman, as well as the multipole moments and the Petrov type. 
  On the other hand, the Hamilton-Jacobi equation is no longer separable, suggesting that the Carter-like constant does not exist in this solution. 
  The standard Newman-Janis algorithm also does not work to construct the new solutions. 
  We also compute the epicyclic frequencies, the location of the innermost stable circular orbits, and the Schwarzschild precession and apply the latter to the recent GRAVITY measurement to place bounds on the deviations away from Kerr-Newman for Sgr A$^*$.

\end{abstract}
\maketitle
%%%%%%%%%%%%%%%%%%%%%%%%%%%%%%%%%%%%%%%%%%%%%%%%%%%%%%%%%%%%%%%%%%%%%%%%%%%%%%%%%%%%%%%%%%%%%%
%%%%%%%%%%%%%%%%%%%%%%%%%%%%%%%%%%%%%%%% INTRODUCTION %%%%%%%%%%%%%%%%%%%%%%%%%%%%%%%%%%%%%%%%
%%%%%%%%%%%%%%%%%%%%%%%%%%%%%%%%%%%%%%%%%%%%%%%%%%%%%%%%%%%%%%%%%%%%%%%%%%%%%%%%%%%%%%%%%%%%%%

\section{Introduction}

General Relativity (GR) has been repeatedly supported by experimental evidence for the past century~\cite{Will:1993ns,Will_SEP,Berti_ModifiedReviewLarge}; a modern example of this success is the advent of gravitational waves, which through the LIGO/Virgo Collaborations has become an incredibly relevant topic of research~\cite{Abbott_IMRcon2,Yunes_ModifiedPhysics,TheLIGOScientific:2016pea,Abbott:2018lct,LIGOScientific:2019fpa,Berti:2018cxi,Berti:2018vdi}. Yet, there are questions in modern physics research which are not modeled completely by Einstein's theory of GR. As such, there are many attempts to define a theory of gravity beyond GR, motivated by, for example, a hopefulness to establish quantum gravity and explain dark energy phenomena~\cite{Berti_ModifiedReviewLarge,Clifton:2011jh,Jain:2010ka,Joyce:2014kja,Koyama:2015vza,DEBook}.

One of the most well-studied  theories beyond GR is the scalar-tensor class, where scalar fields are coupled to gravity. The most generic scalar-tensor theory containing up to 2nd  derivatives in the field equations is Horndeski theory~\cite{Horndeski:1974wa,Deffayet:2011gz}. This theory was later extended to beyond Horndeski theory that contains higher derivatives but avoids ghost modes~\cite{Gleyzes:2014dya}. This theory was further generalized to degenerate higher-order scalar-tensor (DHOST) theories that satisfies a degenerate condition to eliminate Ostrogradski modes. See e.g.~\cite{Kobayashi:2019hrl} for a recent review.   

Another important class of non-GR theories is the vector-tensor class, which generically introduces a preferred direction in spacetime and breaks Lorentz invariance. Such gravitational Lorentz violation is motivated by e.g. ultraviolet completion of gravity~\cite{Horava:2009uw,Blas:2009qj,Blas:2010hb,Blas:2014ira}. Vector-tensor theories also arise within the context of massive gravity~\cite{Gabadadze:2013ria,Ondo:2013wka}. Lorentz violation has been constrained very stringently in the matter sector~\cite{Liberati:2013xla}, while it has not been constrained so strongly in the gravity sector~\cite{Yagi:2013qpa,Yagi:2013ava,Gumrukcuoglu:2017ijh,Oost:2018tcv}. Similar to the scalar-tensor case, generic vector-tensor theories have been constructed within the context of generalized Proca (GP)~\cite{Heisenberg:2014rta,Allys:2015sht,Allys:2016jaq,Rodriguez:2017ckc,De_Felice_2016_2,Allys:2016kbq} and beyond GP theories~\cite{Heisenberg_2016,GallegoCadavid:2019zke}. 
In some cases, there has been evidence of experimental support that GP theories better model cosmological problems than GR~\cite{nakamura2018constraints}. These theories in particular account for vector Galileons which have been used to show applications in current interests of cosmology such as dark matter and dark energy properties~\cite{Felice_2016,Tasinato2014CosmicAF,Rodriguez:2017wkg,De_Felice_2016}. GP and beyond GP theories (together with various other theories) have been constrained from GW170817~\cite{Baker:2017hug,Ezquiaga:2018btd,Domenech_2018}.
These (beyond) GP theories were further generalized to \emph{extended} vector-tensor theories without Ostrogradski modes that satisfy a degenerate condition~\cite{Kimura:2016rzw}.

In this paper, we construct rotating black hole (BH) solutions within a class of extended vector-tensor theories that are valid after GW170817 and study their properties and astrophysical implications. 
Examining the BH solution of alternative theories is of special interest  since the frontiers  of their application are expanding quickly with further data from gravitational-wave experiments~\cite{LIGOScientific:2018mvr}, BH shadow observations with the Event Horizon Telescope~\cite{Akiyama:2019cqa} and stellar motion around Sgr A$^*$ with GRAVITY~\cite{Abuter:2020dou}.
Non-rotating BH solutions within GP frameworks have been constructed in~\cite{Cisterna_2016,Babichev_2017,Heisenberg_2017,Kase_2018}, while rotating \emph{stealth} BH solutions (the Kerr solution with non-trivial vector field configurations) were found in~\cite{Cisterna_2016}. Another rotating BH solution was constructed by applying a disformal transformation to the Kerr-Newman solution~\cite{Filippini:2017kov}. However, a disformal transformation generically changes the propagation speed of perturbations, and such a vector-tensor theory constructed by disformally transforming from Einstein-Maxwell theory has been effectively ruled out from GW170817~\cite{Baker:2017hug,Domenech_2018}. 

Instead, we here apply a generic \emph{conformal} transformation to Kerr-Newman that preserves the causal structure of the original spacetime to derive a rotating BH solution in a subclass of vector-tensor theories valid after GW170817. Such conformal (together with disformal) transformations have widely been used to generate new solutions in non-GR theories (see e.g.~\cite{Bekenstein:1974sf,Yazadjiev:2001bx,Faraoni:2016ozb,Faraoni:2017ecj,Bambi:2016wdn,Chauvineau:2018zjy,BenAchour:2019fdf}). Vector-tensor theories constructed by conformally transforming Einstein-Maxwell theory fall outside of beyond GP theories but are still within the extended vector-tensor theories and are thus healthy and free of Ostrogradski instabilities~\cite{Kimura:2016rzw}.

Let us now briefly summarize our findings. The conformal Kerr-Newman solution preserves the location of singularities, event horizons and ergoregion boundary from Kerr-Newman. We also found that the multipole moments (at least up to the octupole order) and the Petrov type are the same as Kerr-Newman. On the other hand, the separability structure is lost, and the Newman-Janis algorithm does not apply to find a rotating solution from a seed non-rotating one. The epicyclic frequencies and the location of  innermost stable circular orbits (ISCOs) depend on the conformal factor, and thus deviations away from Kerr-Newman may be probed with BH observations of quasi-periodic oscillations and continuum spectrums with X-rays. We also applied the recent Schwarzschild precession measurement of S2 around Sgr A$^*$ with GRAVITY and found bounds on the deviations from Kerr-Newman as a function of the vector charge of Sgr A$^*$.    

The rest of the paper is organized as follows. In Sec.~\ref{sec:theory}, we review the vector-tensor theory that we consider in this paper. In Sec.~\ref{sec:VT}, we explain how we can construct a rotating BH solution. In Sec.~\ref{sec:MetricProperties}, we study various properties of such a BH solution. In Sec.~\ref{sec:NJ}, we investigate whether the Newman-Janis algorithm works in the vector-tensor theory. In Sec.~\ref{sec:astro}, we study astrophysical implications, such as quasi-periodic oscillation frequencies, ISCOs and Schwarzschild precession. We conclude in Sec.~\ref{sec:conclusion} and discuss possible avenues for future work. Throughout, we use the geometric unit of $c=G=1$.

%%%%%%%%%%%%%%%%%%%%%%%%%%%%%%%%%%%%%%%%%%%%%%%%%%%%%%%%%%%%%%%%%%%%%%%%%%%%%%%%%%%%%%%%%%%%%%
%%%%%%%%%%%%%%%%%%%%%%%%%%%%%%%%%%%%% MODIFIED THEORY %%%%%%%%%%%%%%%%%%%%%%%%%%%%%%%%%%%%%%%%
%%%%%%%%%%%%%%%%%%%%%%%%%%%%%%%%%%%%%%%%%%%%%%%%%%%%%%%%%%%%%%%%%%%%%%%%%%%%%%%%%%%%%%%%%%%%%%

\section{Theory}
\label{sec:theory}

Let us first review a healthy vector-tensor theory after GW170817. In~\cite{Kimura:2016rzw}, the authors constructed extended vector-tensor theories that extend (beyond-) GP theories. These theories satisfy degenerate conditions and thus are unaffected by the Ostrogradski instabilities. Moreover, Ref.~\cite{Kimura:2016rzw} shows how conformal/disformal transformation of the metric takes one from one theory to another.

In this paper, we seek to work in a healthy vector-tensor theory after GW170817 that preserves the propagation speed of tensor perturbations to be the speed of light. We start from Einstein-Maxwell theory and apply a certain transformation to the metric. Given that a conformal transformation does not alter the causal structure of a spacetime (and thus has been used widely to e.g. construct Penrose diagrams) whereas disformal transformation changes the propagation speed in general, we only consider the following conformal transformation
\begin{equation}
\label{eq:gbar}
\bar g_{\mu\nu} = \Omega(Y)  g_{\mu\nu} ,
\end{equation}
where we go from an original metric $\bar g_{\mu\nu}$ to a new metric $g_{\mu\nu}$ with $\Omega(Y)$ being an arbitrary function of $Y \equiv A_\mu A^\mu$ with the new vector field $A^\mu$. 
 $Y$ in the new frame and $\bar Y$ ($\equiv\bar A_\mu \bar A^\mu$) in the original frame are related by~\cite{Kimura:2016rzw}
\begin{equation}
\label{eq:Ybar_in_Y}
\bar Y = \frac{Y}{\Omega},
\end{equation}
where we assume that $A_\mu$ is invariant under the transformation: $A_\mu = \bar A_\mu$.

Applying the above conformal transformation to Einstein-Maxwell theory, we arrive at an extended vector-tensor theory with an action~\cite{Kimura:2016rzw}
\begin{align}
\label{eq:action}
S =& \int d^4 x \sqrt{-g} \left\{ \kappa \Omega R - \frac{1}{4} F_{\mu \nu} F^{\mu \nu} + \frac{3 \Omega_Y^2}{4 \Omega}\left[   A^{\mu} A^{\nu} S_{\mu \rho} S_{\nu}{}^{\rho} \right. \right. \nonumber \\
& \left. \left.  +A^{\mu} A^{\nu} F_{\mu \rho} F_{\nu}{}^{\rho}-2 A^{\mu} A^{\nu} F_{\mu}{}^{\rho} S_{\nu \rho} \right] \right\}\,,
\end{align}
with $\kappa = 1/(16\pi)$, $\Omega_Y = d\Omega/dY$ and
\begin{equation}
S_{\mu \nu}=\nabla_{\mu} A_{\nu}+\nabla_{\nu} A_{\mu}, \quad F_{\mu \nu}=\nabla_{\mu} A_{\nu}-\nabla_{\nu} A_{\mu}\,.
\end{equation}
In~\cite{Baker:2017hug,Domenech_2018}, the authors derived  the propagation speed of the tensor modes in GP theories and showed that the coefficient in front of the Ricci scalar in the action needs to be independent of $Y$ (unless we impose a fine-tuned condition among arbitrary functions in the theories) to satisfy the GW170817 bound. This condition does not apply to the action in~\eqref{eq:action} since it does not belong to GP (nor beyond GP) theories (but we stress that it is still a healthy vector-tensor theory).   

\section{Rotating black hole solutions}
\label{sec:VT}

We here apply the conformal transformation to the Kerr-Newman metric, that is a charged, rotating BH solution in Einstein-Maxwell theory, to construct a new rotating BH solution in the vector-tensor theory in Eq.~\eqref{eq:action}. The Kerr-Newman metric is given by~\cite{Newman:1965my}
\begin{eqnarray}\label{eq:KN}
	ds^2_\KN&=& \bar g _{\mu\nu}^\KN dx^\mu dx^\nu \nonumber \\
    &=&-\frac{\Delta}{\rho^2}\left(dt-a \sin^2\theta d\phi\right)^2+\rho^2\left(\frac{dr^2}{\Delta}+d\theta^2\right)\nonumber \\
    && +\frac{\sin^2\theta}{\rho^2}\left[\left(r^2 +a^2\right)d\phi-a \,dt\right]^2, \\
    \label{eq:KN-A}
    \bar A_\mu^\KN &=& \left( -\frac{Qr}{\rho^2},A_r(r),0,\frac{a Q r\sin^2 \theta}{\rho^2} \right),
\end{eqnarray}
with 
\begin{equation}
\Delta=r^2-2Mr+a^2+Q^2, \quad \rho^2=r^2+a^2\cos^2\theta.
\end{equation}
Here $M$ and $Q$ are the BH mass and charge while $a = J/M$ is the Kerr parameter with $J$ representing the spin angular momentum. $A_r(r)$ is an arbitrary function of $r$ due to the gauge symmetry. Although the choice of $A_r$ does not affect the metric in GR, different choices of $A_r$ give different results after the transformation. For simplicity, we choose $A_r = 0.$ $\bar Y^\KN$ is given by
\begin{equation}
\bar Y^\KN=\bar A_\mu^\KN\bar A^\mu_\KN=-\frac{Q^2 r^2}{\Delta  \rho ^2}.
 \end{equation}

We now construct a new BH solution in the vector-tensor theory by conformally transforming the Kerr-Newman solution. 
We introduce the following generic conformal factor:
\begin{equation}\label{eq:Omega}
\Omega (Y) = \left\{1 + f[\bar Y(Y)]  \right\}^{-1}\,,
\end{equation}
for an arbitrary function $f$. The metric reduces to the original one when $f \to 0$. Because the vector-tensor theory metric is, from Eq.~\eqref{eq:gbar},  $g_{\mu\nu}=\Omega^{-1}\bar g_{\mu\nu}^{\KN}$, we choose the form Eq.~\eqref{eq:Omega} for $\Omega$ with exponent $-1$ such that the vector-tensor metric is simply obtained from Kerr-Newman by adding a term that is linear in $f$ times $\bar g^{\KN}_{\mu\nu}$. This solution is fairly simple while keeping $f$ to be general.
The new conformal Kerr-Newman solution is therefore
\begin{equation}
\label{eq:confTrans}
g_{\mu\nu} = \left[1 + f(\bar Y)  \right] \bar g_{\mu\nu}^\KN, \quad A_\mu = \bar A_\mu^\KN.
\end{equation}
To keep the leading asymptotic behavior of the metric at infinity to be the same as Kerr-Newman, we require
\begin{equation}
f(0) = 0.
\end{equation}
Also, we require that $\Omega^{-1}$ is regular and does not vanish everywhere outside the event horizon. We will work on the metric in Eq.~\eqref{eq:confTrans} in most of this paper. 

In Sec.~\ref{sec:astro}, we discuss astrophysical implications of the above conformal Kerr-Newman metric. To put this into context, we will consider a simple example function of
\begin{equation}
\label{eq:f_example}
f(\bar Y) = \frac{\beta}{2} \frac{\bar Y}{1 - \bar Y},
\end{equation}
and the conformal Kerr-Newman solution with this function is given by
\begin{equation}
\label{eq:confTrans2}
g_{\mu\nu} = \left(1-\frac{\beta}{2}\frac{ Q^2 r^2  }{Q^2r^2+\Delta\rho^2}\right) \bar g_{\mu\nu}^\KN. 
\end{equation}
Here, $\beta$ denotes a scale factor of deviation from GR, and $Q$ now corresponds to the vector charge rather than an electric charge. This form of $f(\bar Y)$ in Eq.~\eqref{eq:f_example} with $1-\bar Y$ in the denominator is selected in order to prevent $g_{\mu\nu}$ from diverging at the event horizon since  $\bar Y \to -\infty$ as $\Delta\to 0.$ Notice the denominator of the $\beta$ term in Eq.~\eqref{eq:confTrans2}  never vanishes outside of the event horizon since $Q^2r^2+\Delta\rho^2>0$ everywhere $\Delta\geq 0$, and the above metric simply reduces to $g_{\mu\nu} \to (1-\beta/2)\bar g_{\mu\nu}$ when $\bar Y \to -\infty$. Therefore, the conformal factor is well-behaved and non-vanishing outside of the event horizon when $\beta<2.$

%%%%%%%%%%%%%%%%%%%%%%%%%%%%%%%%%%%%%%%%%%%%%%%%%%%%%%%%%%%%%%%%%%%%%%%%%%%%%%%%%%%%%%%%%%%%%%
%%%%%%%%%%%%%%%%%%%%%%%%%%%%%%%%%%%% METRIC PROPERTIES %%%%%%%%%%%%%%%%%%%%%%%%%%%%%%%%%%%%%%%
%%%%%%%%%%%%%%%%%%%%%%%%%%%%%%%%%%%%%%%%%%%%%%%%%%%%%%%%%%%%%%%%%%%%%%%%%%%%%%%%%%%%%%%%%%%%%%

\section{Metric Properties}
\label{sec:MetricProperties}
 We study several properties of the new BH solution in Eq.~\eqref{eq:confTrans} in this section. Since we require the conformal factor to be regular and non-vanishing everywhere outside the horizon and the conformal transformation does not change the causal structure, the location of the singularity, event horizon and ergoregion boundary is unaltered from Kerr-Newman. The Lorentz signature (the sign of the determinant of the metric) is also unchanged. Below, we focus on finding the multipole moments, separability and Petrov type of the conformal Kerr-Newman solution.

%%%%%%%%%%%%%%%%%%%%%%%%%%%%%%%%%%%%%%%%%%%%%%%%%%%%%%%%%%%%%%%%%%%%%%%%%%%%%%%%%%%%%%%%%%%%%%
%%%%%%%%%%%%%%%%%%%%%%%%%%%%%%%%%%%%%% MULTIPOLE  %%%%%%%%%%%%%%%%%%%%%%%%%%%%%%%%%%%%%%%%%%%%
%%%%%%%%%%%%%%%%%%%%%%%%%%%%%%%%%%%%%%%%%%%%%%%%%%%%%%%%%%%%%%%%%%%%%%%%%%%%%%%%%%%%%%%%%%%%%%

\subsection{Multipole Moments}

 We now derive mass and current multipole moments of the conformal Kerr-Newman solution. We follow Thorne~\cite{RevModPhys.52.299} and compute these quantities in asymptotically Cartesian and mass centered (ACMC) coordinates using the ``flat-space normalized'' basis:
 \begin{equation}
 \bm e_t = \partial_t, \quad 
 \bm e_r = \partial_r, \quad 
 \bm e_\theta =  r^{-1}\partial_\theta, \quad 
 \bm e_\phi = (r \sin\theta)^{-1} \partial_\phi.  
 \end{equation}
 We focus on the asymptotic behavior of the metric at infinity:
 \begin{widetext}
 \begin{eqnarray}
 \label{eq:metric_asympt}
  g_{tt}&=&-1+ \frac{2 M}{r} +\frac{( f'_0-1) Q^2}{r^2}-\frac{ 2 a^2 M \cos^2\theta}{r^3} +\mathcal{O}\left(\frac{1}{r^4}\right),
 \\
 g_{rr}&=&1+\frac{2 M}{r}+\frac{ 4 M^2  - Q^2(1+ f'_0) -  a^2 \sin^2\theta}{r^2} 
 +\frac{2M \left[4 M^2-a^2(2-\cos^2\theta)-2 Q^2(1+ f'_0)\right]}{r^3} 
 +\mathcal{O}\left(\frac{1}{r^4}\right),  \\
 g_{\theta\theta} &=& 1+\frac{a^2 \cos ^2\theta -  Q^2
   f'_0}{r^2}-\frac{2   M Q^2
   f'_0}{r^3}+O\left(\frac{1}{r^4}\right), \\
   g_{\phi\phi} &=& 1+\frac{a^2-  Q^2 f'_0}{r^2}+\frac{2 M (a^2 \sin
   ^2\theta -  Q^2
   f'_0)}{r^3}+O\left(\frac{1}{r^4}\right), \\
g_{t\phi} &=& -\frac{2 a M \sin\theta}{r^2}+\frac{ a Q^2 \sin\theta }{r^3}+\frac{ 2a\sin\theta M( a^2\cos^2\theta+ Q^2 f'_0) }{r^4}+\mathcal{O}\left(\frac{1}{r^5}\right),
 \end{eqnarray}
 where $f'_0 \equiv f'(0)$\footnote{$f_0'=1/2$ when using the metric in Eq.~\eqref{eq:confTrans2}.} and we used $f(0)=0$.
 To move to the ACMC frame, we eliminate terms containing $a^2\sin^2\theta$ and $a^2\cos^2\theta$ at $\mathcal{O}(1/r^2)$ in $g_{rr}$ and $g_{\theta\theta}$ by performing the following coordinate transformation:
\begin{equation}
r=r'+\frac{a^2\cos^2\theta'}{2r'}, \quad
\theta =\theta'- \frac{a^2\cos\theta'\sin\theta'}{2r'^2}, \quad
\phi =\phi', \quad t=t'.
\end{equation}
This yields
\begin{eqnarray}
g_{t't'}&=&-1+ \frac{2 M}{r'} +\frac{Q^2 (  f'_0-1)}{r'^2}-\frac{ 3 a^2 M \cos^2\theta'}{r'^3}
+\mathcal{O}\left(\frac{1}{r'^4} \right),  \\
g_{r'r'}&=&1+\frac{2 M}{r'}+\frac{-a^2 + 4 M^2 - Q^2(1+ f'_0)}{r'^2}+\frac{M[8 M^2-4a^2-4Q^2(1+ f'_0)-a^2\cos^2\theta']}{r'^3} +\mathcal{O}\left(\frac{1}{r'^4} \right), 
\\
g_{\theta'\theta'} &=& 1+{\frac {{a}^{2}- {Q}^{2} f'_0}{{{\it r'}}^{2}}}-{
\frac {2 M{Q}^{2} f'_0}{{{\it r'}}^{3}}}
 + \mathcal{O}\left( \frac{1}{r'^4} \right), \\
 g_{\phi'\phi'} &=& 
 1+{\frac {{a}^{2}-  {Q}^{2} f'_0}{{{\it r'}}^{2}}}+{\frac { 2M ({a}^{2}\sin \theta'^{2}-{Q}^{2} f'_0)}{{{\it r'}}^{3}}}
  + \mathcal{O}\left( \frac{1}{r'^4} \right), 
 \\
g_{t'\phi'}&=&-\frac{2 a M \sin\theta'}{r'^2}+\frac{ a Q^2 \sin\theta' }{r^3}+aM\sin\theta'\frac{5 a^2  \cos^2\theta'+2 Q^2 f'_0}{r'^4}+\mathcal{O}\left(\frac{1}{r'^5} \right), \\
g_{r'\theta'} &=&  -\,{\frac {2M{a}^{2}\sin \theta' \cos \theta'}{{{\it r'}}^{3}}}
+\mathcal{O}\left(\frac{1}{r'^4} \right).
\end{eqnarray}
\end{widetext}
 
We can now compare the above expressions to those in Eq.~(11.4) of~\cite{RevModPhys.52.299}. In particular, from the $1/r'^3$ ($1/r'^4$) term of $g_{t't'}$ ($g_{t'\phi'}$), one can read off the quadrupole mass moment $M_2$ and octupole current moment $S_3$ as
\begin{equation}
 M_2=-8\sqrt{\frac{\pi}{15}}Ma^2, \quad S_{3}=\frac{8}{3}\sqrt{\frac{\pi}{105}}Ma^3, %I_{2m}=0 \mbox{ for }m\neq0,
\end{equation}
which are exactly the same as those for Kerr and Kerr-Newman.
In order to differentiate multipole moments from the Kerr-Newman case, one may need to examine the electromagnetic multipole moments~\cite{Sotiriou:2004ud}, which, to the best of our knowledge, have not been computed within Thorne's formalism.
%%%%%%%%%%%%%%%%%%%%%%%%%%%%%%%%%%%%%%%%%%%%%%%%%%%%%%%%%%%%%%%%%%%%%%%%%%%%%%%%%%%%%%%%%%%%%%
%%%%%%%%%%%%%%%%%%%%%%%%%%%%%%%%%%%% PETROV TYPE  %%%%%%%%%%%%%%%%%%%%%%%%%%%%%%%%%%%%%%%%%%%%
%%%%%%%%%%%%%%%%%%%%%%%%%%%%%%%%%%%%%%%%%%%%%%%%%%%%%%%%%%%%%%%%%%%%%%%%%%%%%%%%%%%%%%%%%%%%%%

\subsection{Petrov Type Classification}

  Petrov types are a way to classify what symmetries the Weyl tensor contains. This is realized by looking at the principal null directions of the Weyl tensor, which will require the null  tetrad formalism. There are six classifications, starting from type I which is the most algebraically general until type O where the Weyl tensor is algebraically special and will vanish~\cite{chandrasekhar1998mathematical}. We note that the Kerr-Newman metric is of Petrov type D. In this section, we show that the conformal Kerr-Newman metric has the same type.
  
To find the Petrov type of a given spacetime, we must find a set of null tetrads $l_\alpha,n_\alpha,m_\alpha,\bar m_\alpha$ (with a bar representing the complex conjugate) such that the following relationships are met~\cite{chandrasekhar1998mathematical}: 
\ba
  g_{\alpha \beta}&=&-l_\alpha n_\beta-n_\alpha l_\beta+m_\alpha\bar{m}_\beta+\bar{m}_\alpha m_\beta, \\
   g^{\alpha \beta}&=&-l^\alpha n^\beta-n^\alpha l^\beta+m^\alpha\bar{m}^\beta+\bar{m}^\alpha m^\beta, 
\ea
with
\ba
  l_\alpha l^\alpha&=&n_\alpha n^\alpha=m_\alpha m^\alpha=\bar{m}_\alpha \bar{m}^\alpha =0, \\
   l_\alpha m^\alpha&=& n_\alpha m^\alpha=l_\alpha \bar{m}^\alpha=n_\alpha \bar{m}^\alpha=0, \\
     l_\alpha n^\alpha&=&-1, \quad m_\alpha \bar{m}^\alpha=1.
\ea
Such null tetrads can be constructed from the orthonormal tetrad {$(e_0){}_{\alpha},(e_1){}_{\alpha},(e_2){}_{\alpha},(e_3){}_{\alpha}$} satisfying
\begin{equation}\label{eq:orthtet}
  g_{\alpha \beta}=-(e_0){}_{\alpha}(e_0){}_{\beta}+(e_1){}_{\alpha}(e_1){}_{\beta}+(e_2){}_{\alpha}(e_2){}_{\beta}+(e_3){}_{\alpha}(e_3){}_{\beta},
\end{equation}
as~\cite{10.1093/mnras/179.3.457}
\ba
  l_\alpha &=& \frac{(e_0){}_{\alpha}+(e_3){}_{\alpha}}{\sqrt{2}}, \quad   n_\alpha=\frac{(e_0){}_{\alpha}-(e_3){}_{\alpha}}{\sqrt{2}}, \\
  m_\alpha&=&\frac{(e_1){}_{\alpha}+i (e_2){}_{\alpha}}{\sqrt{2}}, \quad \bar m_\alpha=\frac{(e_1){}_{\alpha}-i (e_2){}_{\alpha}}{\sqrt{2}}.
\ea

Let us now apply the above formalism to the conformal Kerr-Newman solution. First, the orthonormal tetrads in this metric are related to those for Kerr-Newman (found e.g. in~\cite{DeSmet:2004if}) as
\begin{equation}
(e_\mu){}_\alpha= \Omega^{-1/2} (e_\mu){}_\alpha^\KN.
\end{equation}
From this, we construct the null tetrads that are related to the Kerr-Newman ones as
\begin{align}
l_\alpha &= \Omega^{-1/2} l_\alpha^\KN, \quad n_\alpha = \Omega^{-1/2} n_\alpha^\KN, \\
m_\alpha &= \Omega^{-1/2} m_\alpha^\KN, \quad \bar m_\alpha = \Omega^{-1/2} \bar m_\alpha^\KN.
\end{align}

Next, we determine the Petrov type of the new BH solution by computing the Weyl scalars:~\cite{Stephani:2003tm}
\ba
 \Psi_0 &=&C_{\alpha \beta \gamma \delta}l^\alpha m^\beta l^\gamma m^\delta, \\
 \Psi_1 &=&C_{\alpha \beta \gamma \delta}l^\alpha n^\beta l^\gamma m^\delta, \\
   \Psi_2 &=&C_{\alpha \beta \gamma \delta}l^\alpha m^\beta \bar{m}^\gamma 
  n^\delta, \\
   \Psi_3 &=&C_{\alpha \beta \gamma \delta}l^\alpha n^\beta \bar{m}^\gamma n^\delta, \\
     \Psi_4 &=&C_{\alpha \beta \gamma \delta}n^\alpha \bar{m}^\beta n^\gamma \bar{m}^\delta,
\ea
where $C_{\alpha \beta \gamma \delta}$ is the Weyl tensor. Under a conformal transformation, the Weyl tensor transforms as~\cite{Wald:1984rg}
\begin{equation}
C_{\alpha \beta \gamma \delta} = \Omega^{-1} C_{\alpha \beta \gamma \delta}^\KN.
\end{equation}
Thus, all the Weyl scalars transform in the same way as
\begin{equation}
\Psi_A = \Omega \Psi_A^\KN.
\end{equation}
The conformal Kerr-Newman solution satisfies the relations
\ba\label{eq:IJ}
  I^3=27J^2,\\
  I\neq0\neq J,
\ea
as for Kerr-Newman, where~\cite{Stephani:2003tm}
\ba
 I &=&\Psi_0\Psi_4-4\Psi_1\Psi_3+3\Psi_2^2, \\
   J&=& 
  \begin{vmatrix}
  \Psi_4 & \Psi_3 & \Psi_2 \\ 
  \Psi_3 & \Psi_2 & \Psi_1  \\
  \Psi_2 & \Psi_1 & \Psi_0  \\
  \end{vmatrix} \nonumber \\
 & = &-\Psi_2^3+2\Psi_1\Psi_2\Psi_3+\Psi_0\Psi_2\Psi_4-\Psi_4\Psi_1^2-\Psi_0\Psi_3^2. \nonumber \\
\ea
These relations rule out Types I, III, and N.
Next, we find that the conformal Kerr-Newman solution satisfies
\be\
  K=N=0,
\ee
as for Kerr-Newman, where 
\ba
 K &=&\Psi_1\Psi_4^2-3\Psi_2\Psi_3\Psi_4+2\Psi_3^3, \\
 L &=& \Psi_2\Psi_4-\Psi_3^2,\\
 N &=& 12L^2-\Psi_4^2I.
\ea
This means that the conformal Kerr-Newman metric is of Type D and conformal transformations do not alter Petrov types.

%%%%%%%%%%%%%%%%%%%%%%%%%%%%%%%%%%%%%%%%%%%%%%%%%%%%%%%%%%%%%%%%%%%%%%%%%%%%%%%%%%%%%%%%%%%%%%
%%%%%%%%%%%%%%%%%%%%%%%%%%%%%%%%%%%% HJ SEPARABILITY  %%%%%%%%%%%%%%%%%%%%%%%%%%%%%%%%%%%%%%%%
%%%%%%%%%%%%%%%%%%%%%%%%%%%%%%%%%%%%%%%%%%%%%%%%%%%%%%%%%%%%%%%%%%%%%%%%%%%%%%%%%%%%%%%%%%%%%%

\subsection{Hamilton-Jacobi Equation Separability}

Let us now study whether the Hamilton-Jacobi equation is separable for the vector-tensor BH solution. 
The Hamilton-Jacobi equation is given by~\cite{chandrasekhar1998mathematical}, 
\begin{equation}\label{eq:HJ}
  2\frac{\partial S}{\partial \tau}=g^{\mu \nu}\frac{\partial S}{\partial x^\mu}\frac{\partial S}{\partial x^\nu}.
\end{equation}
Here, $S$ is Hamilton's principal function, and $\tau$ is an affine parameter. 
In order to have a separable solution, Ref.~\cite{chandrasekhar1998mathematical} shows that $S$ must be in the form
\ba\label{eq:S1}
  S=\frac{\delta_1 \ \tau}{2}-E \ t+L_z  \ \phi+F_r(r)+F_\theta(\theta),\\\nonumber
\ea
where $E$ is specific energy, $L_z$ is specific angular momentum, and $F_r$ and $F_\theta$ are functions of $r$ and $\theta$, respectively. Additionally, note $\delta_1$ takes the form
\be
\delta_1=
\begin{cases}
0 &\text{for null geodesics},\\
1 &\text{for timelike geodesics}.
\end{cases}
\ee
If $S$ is separable, we can find the Carter-like constant of motion which indicates that we can present the solution of the geodesic equation as integral expressions explicitly.

Substituting Eq.~\eqref{eq:S1} into Eq.~\eqref{eq:HJ}, one finds 
\bw
\be
  \label{eq:S3}
\Delta\left(\frac{\partial F_r}{\partial r}\right)^2+ \left(\frac{\partial F_\theta}{\partial \theta}\right)^2+\csc^2 \theta \left(a\sin^2\theta E-L_z\right)^2-\frac{\left[a L_z-(a^2+r^2)E\right]^2}{\Delta}-\delta_1\rho^2 \left[1+ f(\bar Y)\right]=0.
\ee
\ew

Now, we can see that the null geodesic equation has a separable solution; when we take $\delta_1=0$ in Eq.~\eqref{eq:S3}, no individual term has both $r$ and $\theta$ dependence. This is consistent with the results found by Walker and Penrose in~\cite{walker1970} which states that all Petrov type D solutions have separable solutions to the Hamilton-Jacobi equation for null case. We can see that Eq.~\eqref{eq:S3} is the same as the Kerr-Newman case since $f$ is absent when $\delta_1=0.$

Next, we examine the separability of the original and conformal Kerr-Newman solution based on the above equation in the timelike case.
When $f = 0$, the coefficient of $\delta_1$ can be separated to functions of either $r$ or $\theta$. Thus each term in Eq.~\eqref{eq:S3} is a function of either $r$ or $\theta$ (or a constant), making the equation separable. On the other hand, when $f \neq 0$, the $f$-dependent term is a function of both $r$ and $\theta$ in general. Therefore, we conclude that the Hamilton-Jacobi equation for the conformal Kerr-Newman solution in the Boyer-Lindquist-like coordinates is not separable for the timelike case, which suggests that a Carter-like constant may not exist in general. 

To elaborate on this point further, we perform another test on the separability structure of the conformal Kerr-Newman solution. Benenti and Francaviglia~\cite{Benenti} showed that if a spacetime admits a separable structure, its metric components can be expressed in the following form (see~\cite{Vigeland:2011ji,Johannsen:2015pca,Konoplya:2018arm,Papadopoulos:2018nvd,Carson:2020dez} for related works):
\begin{align}
\label{eq:Benenti}
g^{rr} =& \frac{\bar Q(r)}{r^2 + p^2}\,, \quad g^{\theta\theta} = \frac{\bar P(p)}{(r^2 + p^2) a^2 \sin^2 \theta}\,, \nonumber \\
g^{AB} = &\frac{\bar Q(r)}{r^2 + p^2}\zeta_r^{AB}(r) + \frac{\bar P(p)}{r^2 + p^2}\zeta_p^{AB}(p)\,, 
\end{align}
where $p \equiv a \cos\theta$ and $(A,B)$ are either $t$ or $\phi$. $\bar Q(r)$ and $\zeta_r^{AB}(r)$ are arbitrary functions of $r$ while $\bar P(p)$ and $\zeta_p^{AB}(p)$ are arbitrary functions of $p$. We checked that the conformal Kerr-Newman in the Boyer-Lindquist-like coordinates cannot be mapped to the above form. For example, $g^{rr}$ and $g^{\theta\theta}$ are
\begin{align}
g^{rr} =& \frac{1}{r^2 + p^2} \frac{r^2 - 2 M r + a^2 + Q^2}{1 +  f(\bar Y)}\,,  \\
g^{\theta\theta} =& \frac{1}{r^2 + p^2} \frac{1}{1 +  f(\bar Y)}\,,
\end{align}
and thus they cannot be mapped to Eq.~\eqref{eq:Benenti} unless $f$ is a constant. There is a possibility that one needs to perform a coordinate transformation to map the conformal Kerr-Newman solution to Eq.~\eqref{eq:Benenti}. However, if we perform such a transformation between $(r,\theta)$, it generates $g^{r\theta}$ which is absent in Eq.~\eqref{eq:Benenti}. Therefore, we believe the conformal Kerr-Newman cannot be mapped Eq.~\eqref{eq:Benenti} in any coordinates.

%%%%%%%%%%%%%%%%%%%%%%%%%%%%%%%%%%%%%%%%%%%%%%%%%%%%%%%%%%%%%%%%%%%%%%%%%%%%%%%%%%%%%%%%%%%%%%
%%%%%%%%%%%%%%%%%%%%%%%%%%%%%%%%%%%% NEWMAN JANIS  %%%%%%%%%%%%%%%%%%%%%%%%%%%%%%%%%%%%%%%%%%%
%%%%%%%%%%%%%%%%%%%%%%%%%%%%%%%%%%%%%%%%%%%%%%%%%%%%%%%%%%%%%%%%%%%%%%%%%%%%%%%%%%%%%%%%%%%%%%

\section{Newman-Janis Algorithm}
\label{sec:NJ}

We next study whether the Newman-Janis algorithm~\cite{Newman:1965tw,Drake:1998gf,Erbin:2016lzq} applies to the conformal Kerr-Newman solution in the vector-tensor theory. The algorithm was developed to construct a rotating BH solution from a static one. Such algorithm works for constructing the Kerr and Kerr-Newman solutions, though it does not necessarily hold in theories beyond GR~\cite{Hansen:2013owa}\footnote{See e.g.~\cite{Yazadjiev:1999ce,Wei:2020ght,Kumar:2020owy} for examples in which the Newman-Janis algorithm works in non-GR theories.}. We will primarily follow the prescription by Giampieri~\cite{Erbin:2014aya} that is simpler than the original algorithm by Newman and Janis.  

 \subsection{Algorithm}
 
 Let us first prescribe the algorithm following~\cite{Erbin:2016lzq}. 
 
 \begin{enumerate}
 
  \item \emph{Seed metric}
  
   The algorithm begins with the seed metric and gauge field for a non-rotating BH spacetime, given by
  \begin{eqnarray}
   ds^2&=&-f_t(r)dt^2+f_r(r) dr^2+f_\Omega (r) d\Omega^2, \\
   %d\Omega^2&=&d\theta^2+\sin^2\theta d\phi^2, \\
   A&=&\varphi(r)dt.
   \end{eqnarray}
   \item \emph{Null coordinate transformation}
   
    Next, the metric and gauge field are transformed into null coordinates $(u,r,\theta,\phi)$ where 
   \begin{equation}
     dt=du-\sqrt{\frac{f_r}{f_t}}dr,
     \end{equation}
   which yields  
   \begin{eqnarray}
     ds^2 &=& -f_tdu^2+2\sqrt{f_tf_r}drdu+f_\Omega d\Omega^2, \\
     A &=&\varphi \left(du-\sqrt{\frac{f_r}{f_t}}dr \right).
     \end{eqnarray}
   One can find a gauge transformation such that the $r$ component of  $A$ vanishes. Doing so, one finds 
        \begin{equation}
        A=\varphi du.
        \end{equation}

   \item \emph{Giampieri prescription}
   
   The next step is to introduce the transformation  
 \begin{eqnarray}
 \label{eq:u_transform}
   du&=&du'-i a \sin\theta \ d\theta, \\
   \label{eq:r_transform}
 dr&=&dr'+i a \sin\theta \ d\theta. 
 \end{eqnarray}
  One then performs further transformation
  \begin{equation}\label{eq:Giampieri}
    id\theta=\sin\theta d\phi,
  \end{equation} as an alternative to the Newman-Janis prescription. This prescription is simpler as it avoids the use of Newman-Penrose formalism. One then finds
   \begin{eqnarray}
     du&=&du'-a\sin^2\theta d\phi, \\
      dr&=&dr'+a\sin^2\theta d\phi.
   \end{eqnarray} 
   
   \item \emph{Radial transformation}

One then needs to introduce a complex radial coordinate with the following rules:
   \begin{eqnarray}\label{eq:comp1}
     r &\rightarrow & \frac{r+\bar{r}}{2}=\Re(r), \\
     \label{eq:comp2}
     \frac{1}{r} &\rightarrow &\frac{1}{2}\left(\frac{1}{r}+\frac{1}{\bar{r}}\right)=\frac{\Re(r)}{|r|^2}, \\
\label{eq:comp3}
     r^2 & \rightarrow & |r|^2.
   \end{eqnarray} 
   However, Eq.~\eqref{eq:r_transform} implies 
   \begin{equation}\label{eq:comp4}
   |r|^2=\rho^2,  \quad \Re(r)=r.
   \end{equation} 
Thus, $r$ remains unchanged while $1/r$ and $r^2$ transform as
\ba 
\label{eq:oneoverr}
\frac{1}{r} &\rightarrow &\frac{r}{\rho^2}, \\
\label{eq:rsquared}
     r^2 & \rightarrow & \rho^2.
\ea
 The metric and gauge field are given by
   \begin{eqnarray}
    ds^2 &=& -\bar{f}_t\left(du'+\sqrt{\frac{\bar{f}_r}{\bar{f}_t}} dr'+\omega \sin\theta d\phi\right)^2\nonumber\\
    &&+2a\bar{f}_r\sin^2\theta drd\phi +\bar{f}_\Omega(d\theta^2+\sigma\sin^2\theta d\phi^2), \\
    A &=& \bar{\varphi}(du'+a\sin^2\theta d\phi),
    \end{eqnarray} 
   where 
   \begin{eqnarray}
   \omega &=& a\sin\theta\left(-1+\sqrt{\frac{\bar{f}_r}{\bar{f}_t}}\right), \\
   \sigma &=& 1+\frac{\bar{f}_r}{\bar{f}_\Omega}a^2\sin^2\theta.
   \end{eqnarray} 
A bar on $f$ and $\varphi$ refers to these functions after the radial transformation in Eqs.~\eqref{eq:oneoverr} and~\eqref{eq:rsquared}.

   \item \emph{Boyer-Lindquist-like Coordinates}

    The final step of the algorithm is to transform into Boyer-Lindquist-like coordinates which describe the metric with the minimal number of components. In order to eliminate the $(t,r)$ and $(r,\phi)$ components of the metric, we perform the following coordinate transformation\footnote{We omit the prime on $r$.}:  
   \begin{equation}
   du'=dt'-g(r)dr, \quad   d\phi=d\phi'-h(r)dr.
   \end{equation} 
  Here 
  \begin{equation}
  g(r)=\frac{\sqrt{(\bar{f}_t\bar{f}_r)^{-1}}\bar{f}_\Omega+a^2\sin^2\theta}{\bar \Delta}, \quad h(r)=\frac{a}{\bar \Delta},
  \end{equation}
  with $\bar \Delta(r)=\bar{f}_\Omega/\bar{f}_r+a^2\sin^2\theta.$ This transformation is possible only if $g$ and $h$ are functions of $r$ only. 
  After carrying out this transformation, we arrive at the final metric and gauge field expression (omitting primes on $t$ and $\phi$): 
  \begin{eqnarray}
  \label{eq:NJ-final}
    ds^2&=& -\bar{f}_t(dt+\omega \sin\theta d\phi)^2+\frac{\bar{f}_\Omega}{\bar \Delta}dr^2 \nonumber \\
  &&  +\bar{f}_\Omega(d\theta^2+\sigma\sin^2\theta d\phi^2), \\
  \label{eq:NJ-final2}
  A &=&\bar{\varphi}\left(dt-\frac{\bar{f}_\Omega}{\bar\Delta\sqrt{\bar{f}_t\bar{f}_r}}dr-a\sin^2\theta d\phi \right).
  \end{eqnarray} 
Again, the $r$ component of the gauge field can usually be eliminated upon an appropriate gauge transformation.
  \end{enumerate}

 \subsection{Application to BH Solutions }

\subsubsection{Kerr-Newman}

As an example, let us see how the above algorithm can be applied to derive the Kerr-Newman metric. The seed metric is Reissner-Nordstr\"{o}m, whose seed functions are given by
\begin{eqnarray}\label{eq:seedRN}
f_t &=& f_r^{-1} = 1-\frac{2M}{r}+\frac{Q^2}{r^2}, \\
 f_\Omega(r)&=&r^2, \quad \varphi = -\frac{Q}{r}.
\end{eqnarray}
By complexifying the radial coordinate, these functions can be turned into
\begin{eqnarray}\label{eq:seedRN2}
\bar f_t &=& \bar f_r^{-1} = 1-\frac{2Mr}{\rho^2}+\frac{Q^2}{\rho^2}, \\
 \bar f_\Omega(r)&=&\rho^2, \quad \bar \varphi = -\frac{Qr}{\rho^2}.
\end{eqnarray}
Plugging in these functions into Eqs.~\eqref{eq:NJ-final} and~\eqref{eq:NJ-final2}, one can correctly reproduce the Kerr-Newman metric in Eq.~\eqref{eq:KN} and its associated gauge field in Eq.~\eqref{eq:KN-A}.

\subsubsection{Conformal Kerr-Newman}

Let us next see whether one can use the algorithm to derive the conformal Kerr-Newman metric in Eq.~\eqref{eq:confTrans2} in the vector-tensor theory. Given that the Kerr-Newman solution can be reproduced via the algorithm, the question reduces to whether the conformal factor is correctly recovered under the algorithm. 

Since such a factor depends only on $\bar Y$, the question further reduces to whether $\bar Y$ is correctly recovered through the algorithm. $\bar Y$ in the Reissner-Nordstr\"{o}m BH solution is given by
\begin{equation}
\bar Y|_{a=0} = - \frac{Q^2}{r^2} \left( 1 - \frac{2M}{r} + \frac{Q^2}{r^2} \right)^{-1}. 
\end{equation}
Using the replacement in Eqs.~\eqref{eq:oneoverr} and~\eqref{eq:rsquared}, this changes to
\begin{align}
\bar Y|_{a=0} \to & - \frac{Q^2}{\rho^2} \left( 1 - \frac{2Mr}{\rho^2} + \frac{Q^2}{\rho^2} \right)^{-1} \nonumber \\
=& - \frac{Q^2}{\rho^2 - 2M r + Q^2} \nonumber \\
\neq & \bar Y.
\end{align}
Thus, the Newman-Janis algorithm does not work for the conformal Kerr-Newman BH solution in the vector-tensor theory.
\begin{figure*}[t]
  \centering
   \includegraphics[width=8.5cm]{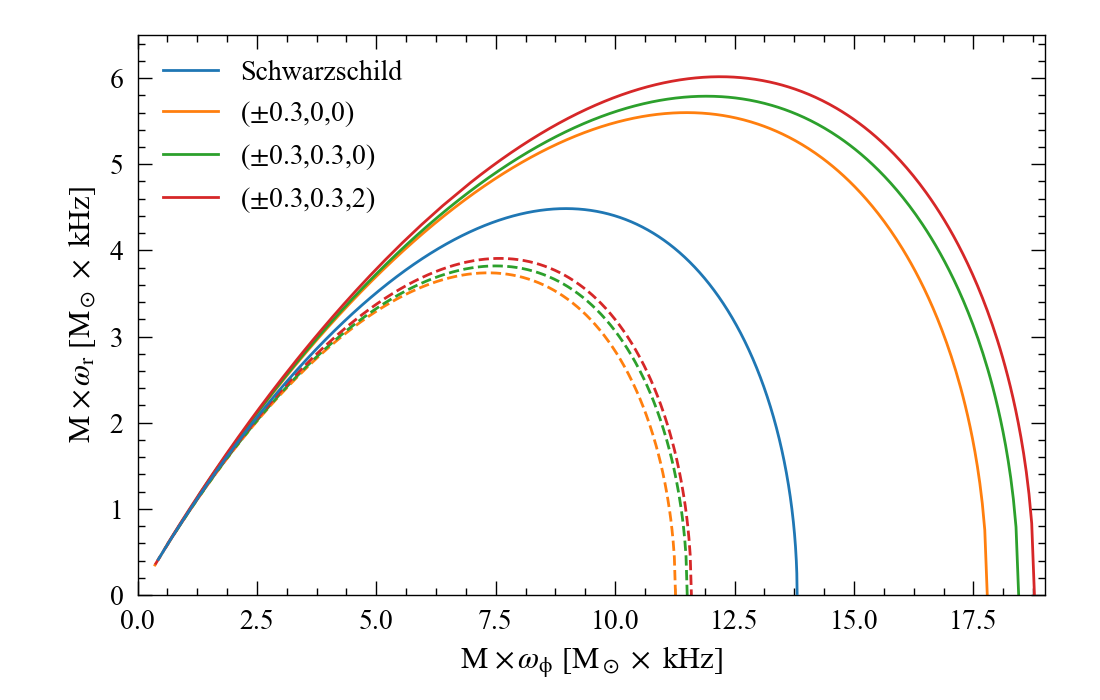}
  \includegraphics[width=8.5cm]{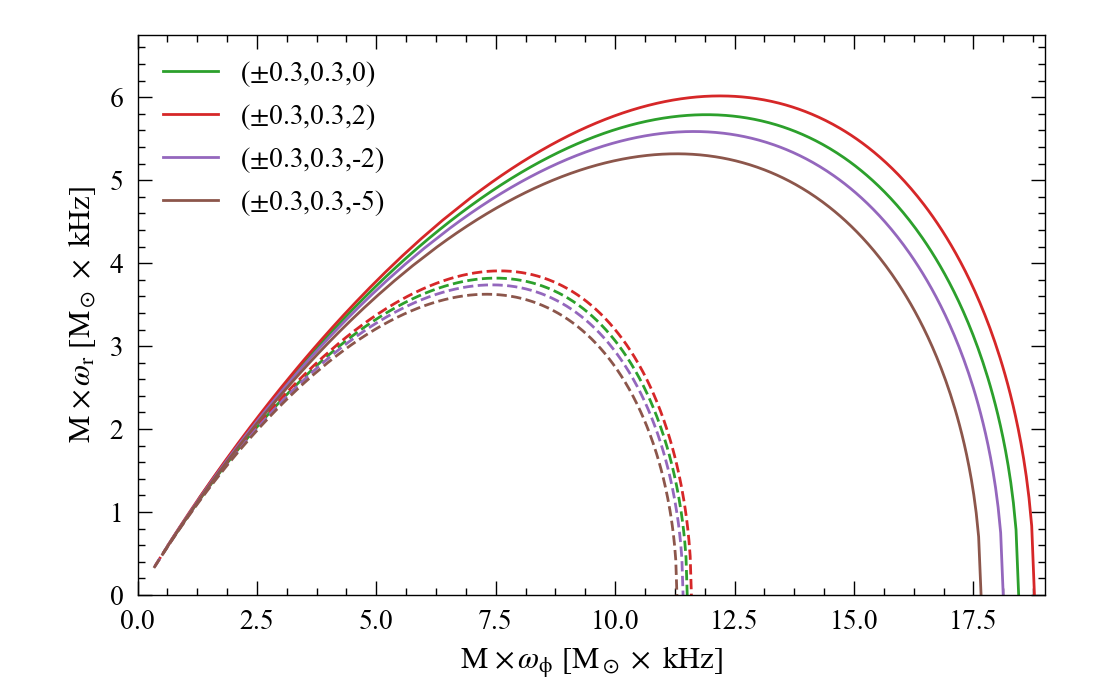}
    \caption{(left) Epicyclic frequency versus orbital frequency for various ($a/M$, $Q/M$, $\beta$).  The solid (dashed) lines correspond to the positive (negative) spin parameters.  (right) Similar to the left plot; however we instead show how variation of $\beta$ affects the outcome of frequencies. Note that $\beta=2$ results in $g_{\mu\nu} = 0$ at the event horizon, but we include it here in order to show maximum deviation with positive $\beta$.}
   \label{fig:QPO}
\end{figure*}

%%%%%%%%%%%%%%%%%%
\section{Astrophysical Applications}
\label{sec:astro}

In addition to the properties already presented with the new metric (see Sec.~\ref{sec:MetricProperties}), we may investigate some applications which can be studied in an astrophysical sense.  Because the vector-tensor BH solution has the same Hamilton-Jacobi equation as the Kerr-Newman metric for null geodesics, it would be difficult to test test this theory using BH shadow observations recently made with the Event Horizon Telescope~\cite{Akiyama:2019cqa}. Thus, we shift our focus to timelike particle motion.
In particular, we consider epicyclic frequencies and ISCOs of particles orbiting around a black hole and the Schwarzschild precession of S2 around Sgr A$^*$. 
We assume that both particles and S2 are uncharged and thus follow geodesic motions.

\subsection{Epicyclic Frequencies and ISCOs }

A natural starting point to investigate is that of geodesic motion, which can be studied by investigating the consequences that our metric imposes on timelike particles. In this subsection, we consider an example conformal function in Eq.~\eqref{eq:f_example} and work in a conformal Kerr-Newman solution in Eq.~\eqref{eq:confTrans2}.

Due to the static and axisymmetric nature of our spacetime, we are immediately allowed to define two conserved quantities which we will call specific energy ($E$) and specific angular momentum ($L$) based on the timelike ($\xi^{(t)}$) and azimuthal ($\xi^{(\phi)}$) Killing vectors of our spacetime.  These quantities are defined by
\begin{eqnarray}
E = -u^\alpha \xi^{(t)}_\alpha, \quad L = u^\alpha \xi^{(\phi)}_\alpha,
\end{eqnarray}
where $u^\alpha = dx^\alpha/d\tau$ with $\tau$ representing the proper time.
The normalization of the four velocity of a timelike particle indicates
\begin{equation}
    g_{\alpha \beta}u^\alpha u^\beta = -1.
\end{equation}
From this expression, and taking the case of equatorial geodesics ($\theta=\pi/2$), we obtain
\begin{equation}
\label{eq:PotentialEqn}
    \frac{1}{2}\dot{r}^2 = V_\mathrm{eff}(r),
\end{equation}
where the dot represents a derivative with respect to the proper time while $V_\mathrm{eff}(r)$ is the effective potential for our system.

From the effective potential we may study the motion of test particles for a circular orbit. This is found by observing the conditions $V_\mathrm{eff}(r)=0$ and $\partial_r V_\mathrm{eff}(r)=0$.  We define the angular velocity relative to a rest frame at spatial infinity by
\begin{equation}
    \omega_{\phi} \equiv \frac{u^\phi}{u^t}\,,
\end{equation}
which can be rewritten in terms of our conserved quantities as
\begin{equation}
\label{eq:OrbitalFreq}
    \omega_\phi =- \frac{E\,g_{t\phi}+L\,g_{tt}}{E\,g_{\phi\phi}+L\,g_{t\phi}}.
\end{equation}
In addition to this, there is another frequency which is of interest to study.  That is the epicyclic frequency, defined to be\footnote{Note that this is different than the definition presented in~\cite{Maselli:2014fca,Glampedakis:2016pes}.  However, this stems from differences in defining the effective potential term, and does not contribute differences in the calculated value of $\omega_r$.}
\begin{equation}
\label{eq:EpiFreq}
    \omega_r^2 = -\frac{1}{\dot{t}^2}\frac{\partial^2\,V_\mathrm{eff}(r)}{\partial r^2}\,.
\end{equation}
Equation~(\ref{eq:EpiFreq}) is found by perturbing Eq.~\eqref{eq:PotentialEqn} about the ISCO, which we define to be the radius solving the equation
\begin{equation}
  \frac{dE}{dr}\Bigg|_{r=R_{\ISCO}}=0\,.
\end{equation}
Figure~\ref{fig:QPO} presents one frequency against another for various combinations of $a$, $Q$ and $\beta$. 
Note that the values are chosen to emphasize the difference between the vector-tensor theory and GR; however, most of the values presented are reasonable within the tests we describe in Sec.~\ref{ssec:SchwarzPrecession} and the restrictions to $\beta$ that we have already outlined in Sec.~\ref{sec:VT}. Although $\beta=2$ leads to $g_{\mu\nu}=0$ at the event horizon, we include this to show maximum deviation. We also note that the bound on the BH electric charge in GR from the binary BH merger event GW150914 is not so stringent,  $Q/M < 0.4$~\cite{bozzola2020general}. Moreover, $Q$ is a parameter that is unique to each BH. Hence, even if $Q/M$ has been constrained stringently with one BH, it does not mean that other BHs need to have small charges.

\begin{figure}[t]
  \centering
\begin{subfigure}[b]{1\columnwidth}
  \includegraphics[width=1\linewidth]{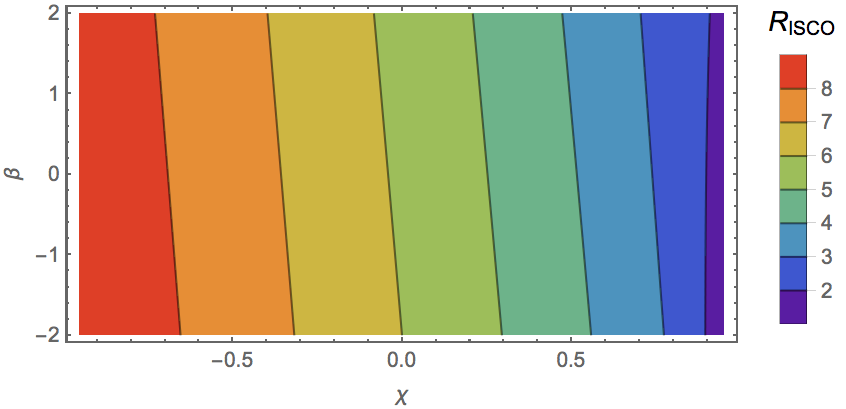}
\end{subfigure}

\begin{subfigure}[b]{1\columnwidth}
  \includegraphics[width=1\linewidth]{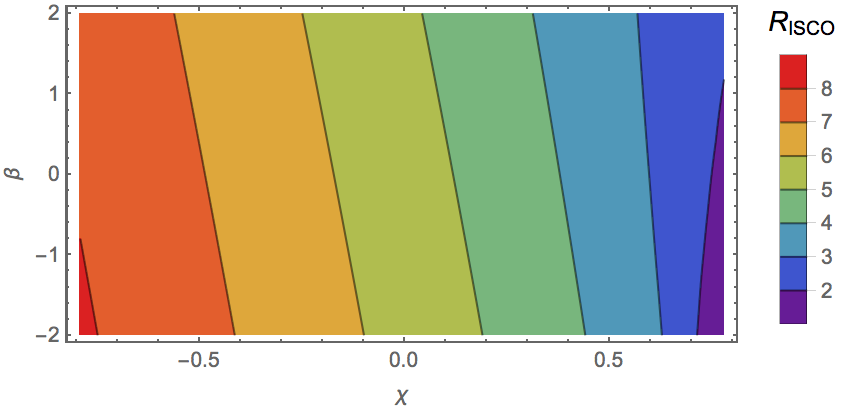}
\end{subfigure}

\begin{subfigure}[b]{1\columnwidth}
  \includegraphics[width=1\linewidth]{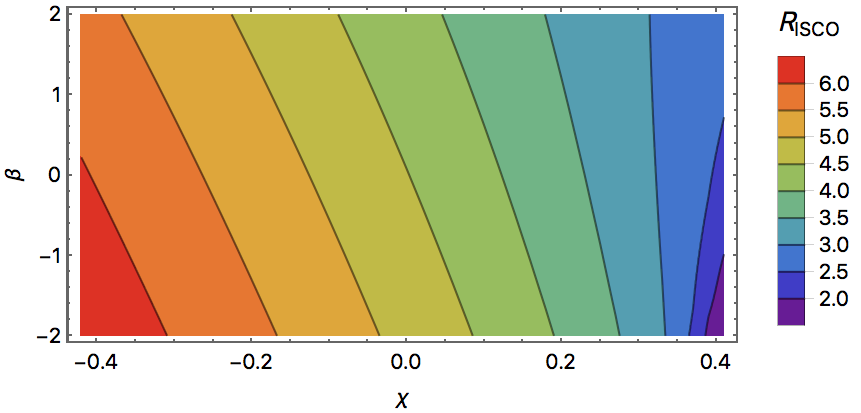}
\end{subfigure}

    \caption{Various ISCO locations in units of $M$ for $Q/M=0.3$ (top), $Q/M=0.6$ (middle), and $Q/M=0.9$ (bottom) as functions of the dimensionless spin $\chi = a/M$ and $\beta$.  We bound the spins to values which admit real values for the event horizon.}
  \label{fig:ISCO}
\end{figure}

We also show how the ISCO is modified for various values of spin, charge, and $\beta$ in Fig.~\ref{fig:ISCO}. Observe that the ISCO locations (considered to be the inner edge of typical accretion disks and one may  extract these from BH observations with X-rays) varies from the GR case ($\beta=0$) especially when the vector charge is large. 

We may apply Eqs.~\eqref{eq:OrbitalFreq} and~\eqref{eq:EpiFreq} to study a phenomena known as quasi-periodic oscillations (QPOs).  
The origin of QPOs is still an open area of discussion, with proposed explanations ranging from orbital resonances~\cite{Abramowicz:2001bi} to
the motion of relativistic matter near a compact object~\cite{Stella:1998mq}.  
Regardless of the explanation, observations of QPOs could be used to test GR and provide insights into the nature of gravity close to compact objects~\cite{Glampedakis:2016pes,Saffer:2019hqn}.  
The fact that there is a noticeable difference in various curves in Fig.~\ref{fig:QPO} suggests that one can in principle use QPO observations to probe the vector-tensor theory once the systematic errors are under control.  
It should be noted that different combinations of parameters have the potential to yield similar curves (e.g. the negative spin curves in Fig.~\ref{fig:QPO}).  Therefore, this method may only be useful in placing limits on the combination of parameters, and other methods of constraint will be needed to break this degeneracy.

\subsection{Schwarzschild Precession} \label{ssec:SchwarzPrecession}

Another astrophysical test of BHs is to use the  Schwarzschild precession. Recently, GRAVITY measured the Schwarzschild precession of a star S2 orbiting around Sgr A$^*$~\cite{Abuter:2020dou}. The Schwarzschild precession is given by
\begin{equation}
\Delta \omega = \frac{6\pi M}{\bar a (1-e^2)} f_\mathrm{sp},
\end{equation}
where $\bar a$ is the semi-major axis while $e$ is the orbital eccentricity. $f_\mathrm{sp}$ is a parameter controlling the relativistic effect. GR predicts $f_\mathrm{sp} = 1$ while $f_\mathrm{sp} \to 0$ corresponds to the Newtonian limit. GRAVITY placed a bound on this parameter as $f_\mathrm{sp} = 1.1 \pm 0.19$~\cite{Abuter:2020dou}. 
$f_\mathrm{sp}$ is also related to the parameterized post-Newtonian (PPN) parameters as~\cite{Will:1993ns}
\begin{equation}
f_\mathrm{sp} = \frac{2-2\gamma_\PPN - \beta_\PPN}{3}.
\end{equation}
These PPN parameters enters in the metric as
\begin{align}
\label{eq:metric_PPN}
g_{tt} &= -1 + \frac{2M}{r} + 2(\beta_\PPN - \gamma_\PPN)\frac{M^2}{r^2} + \mathcal{O}\left(\frac{M^3}{r^3}\right), \\
g_{rr} &= 1 + 2\gamma_\PPN \frac{M}{r} +\mathcal{O}\left(\frac{M^2}{r^2}\right). 
\end{align}
GR is recovered in the limit $\gamma_\PPN \to 1$ and $\beta_\PPN \to 1$. 

\begin{figure}[h]
  \centering
  \includegraphics[width=8.5cm]{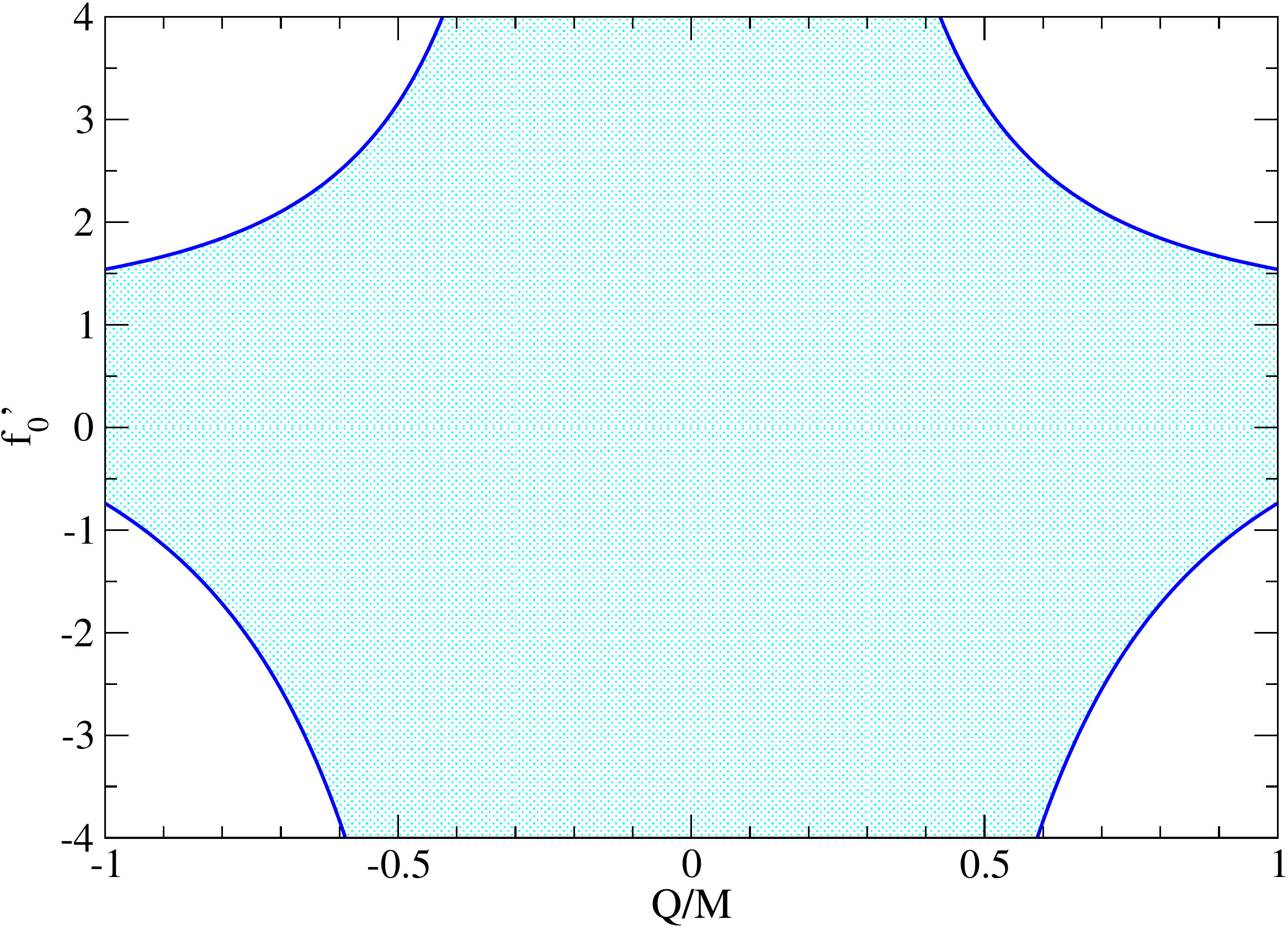}
    \caption{Bounds on the theoretical parameter $f_0'$ as a function of the dimensionless vector charge $Q/M$ of Sgr A$^*$ from the Schwarzschild precession measurement by GRAVITY. The shaded region is the allowed region.}
   \label{fig:precession}
\end{figure}

We can use the above Schwarzschild precession measurement to constrain the vector-tensor theory. We assume that the vector charge of S2 is negligible (so that a Coulomb-like interaction force is absent) and that it is in a geodesic motion as in GR. Comparing Eqs.~\eqref{eq:metric_asympt} and~\eqref{eq:metric_PPN}, we find
\begin{equation}
\label{eq:PPN_mapping}
\gamma_\PPN = 1, \quad \beta_\PPN -1 = \frac{Q^2}{2M^2} (\beta f_0' -1)\,.
\end{equation}
When $\gamma_\PPN = 1$, GRAVITY's measurement can be mapped to bounds on the PPN parameters as
\begin{equation}
\label{eq:beta_bound}
\beta_\PPN - 1 = -0.3 \pm 0.57.
\end{equation}
A similar bound has been obtained in~\cite{Gainutdinov:2020bbv}. From Eqs.~\eqref{eq:PPN_mapping} and~\eqref{eq:beta_bound}, one can constrain the parameter space of the vector-tensor theory. Figure~\ref{fig:precession} shows the bound on $f_0'$ as a function of the vector charge $Q/M$ for Sgr A$^*$\footnote{Note that most values presented in Fig.~\ref{fig:QPO} and~\ref{fig:ISCO} (with $f_0' = \beta/2$) are mostly within these derived bounds. Moreover, there is no problem in choosing parameters outside of the allowed region in Fig.~\ref{fig:precession} since $Q/M$ in the figure is specific to Sgr A* and if its charge is close to 0, $f_0'$ remains almost unconstrained.}. As expected, we cannot place any bounds when $Q/M=0$, while the bound becomes stronger for larger $|Q/M|$. In addition, the figure also shows that in GR ($\beta = 0$), the GRAVITY measurement does not constrain the electric charge of Sgr A$^*$.

%%%%%%%%%%%%%%%%%%%
\section{Conclusions}
\label{sec:conclusion}

We derived rotating BH solutions in a certain class of extended vector-tensor theory by applying a conformal transformation to the Kerr-Newman solution in GR. Such transformation does not alter the causal structure of spacetime and thus the theory is healthy and valid even after GW170817. 

We then studied various properties of the conformal Kerr-Newman solution by keeping the conformal factor arbitrary (but require it to be regular and non-vanishing outside the event horizon and asymptotes to 1 at infinity). We found that the locations of the singularity, event horizons, and ergosphere are unaffected from the Kerr-Newman case. We also found that the multipolar structure of the metric is the same as that of Kerr-Newman (though the asymptotic behavior of the metric at infinity acquires corrections from GR), while the metric does not allow for the separability of the Hamilton-Jacobi equation, suggesting that a Carter-like constant does not exist in such a solution. The new BH solution can be  classified as Petrov type D, which is identical to Kerr-Newman since the conformal transformation does not alter the Petrov type. We also checked that the standard Newman-Janis algorithm cannot be applied to derive the solution. 

We finally studied astrophysical implications, such as epicyclic frequencies and ISCOs that are related to quasi-periodic oscillations and inner edges of accretion disks. We also used the recent Schwarzschild precession measurement of S2 orbiting around Sgr A$^*$ using GRAVITY to place bounds on a theoretical parameter as a function of the vector charge of Sgr A$^*$.  

Various avenues exist for future work. For example, one can consider perturbations of the new BH solution to study its stability and derive quasi-normal mode ringdown frequencies and damping times (see~\cite{Kokkotas:1993ef,Berti:2005eb,Pani:2013ija,Pani:2013wsa,Mark:2014aja,Dias:2015wqa,Giorgi:2020ujd} for related works on Kerr-Newman). One can also 
study how gravitational waveforms from binary BH coalescences are modified from GR in this theory. 
It is interesting to extend the astrophysical applications presented in this paper by revisiting e.g. iron line spectrums~\cite{Yang:2018wye} in vector-tensor theories. It would also be interesting to perform gravitational collapse simulations in this theory to study if there is any expected range of $Q$ in this theory.

%%%%%%%%%%%
\acknowledgments

We thank Atsushi Naruko and George Pappas for helpful discussions on the extended vector-tensor theories, separability structure and the Thorne multipole moments.
K.Y. acknowledges support from NSF Award PHY-1806776, NASA Grant 80NSSC20K0523, a Sloan Foundation Research Fellowship and the Ed Owens Fund. 
K.Y. would like to also acknowledge support by the COST Action GWverse CA16104 and JSPS KAKENHI Grants No. JP17H06358.

\bibliography{bibliography.bib}

\begin{thebibliography}{100}%
\makeatletter
\providecommand \@ifxundefined [1]{%
 \@ifx{#1\undefined}
}%
\providecommand \@ifnum [1]{%
 \ifnum #1\expandafter \@firstoftwo
 \else \expandafter \@secondoftwo
 \fi
}%
\providecommand \@ifx [1]{%
 \ifx #1\expandafter \@firstoftwo
 \else \expandafter \@secondoftwo
 \fi
}%
\providecommand \natexlab [1]{#1}%
\providecommand \enquote  [1]{``#1''}%
\providecommand \bibnamefont  [1]{#1}%
\providecommand \bibfnamefont [1]{#1}%
\providecommand \citenamefont [1]{#1}%
\providecommand \href@noop [0]{\@secondoftwo}%
\providecommand \href [0]{\begingroup \@sanitize@url \@href}%
\providecommand \@href[1]{\@@startlink{#1}\@@href}%
\providecommand \@@href[1]{\endgroup#1\@@endlink}%
\providecommand \@sanitize@url [0]{\catcode `\\12\catcode `\$12\catcode
  `\&12\catcode `\#12\catcode `\^12\catcode `\_12\catcode `\%12\relax}%
\providecommand \@@startlink[1]{}%
\providecommand \@@endlink[0]{}%
\providecommand \url  [0]{\begingroup\@sanitize@url \@url }%
\providecommand \@url [1]{\endgroup\@href {#1}{\urlprefix }}%
\providecommand \urlprefix  [0]{URL }%
\providecommand \Eprint [0]{\href }%
\providecommand \doibase [0]{http://dx.doi.org/}%
\providecommand \selectlanguage [0]{\@gobble}%
\providecommand \bibinfo  [0]{\@secondoftwo}%
\providecommand \bibfield  [0]{\@secondoftwo}%
\providecommand \translation [1]{[#1]}%
\providecommand \BibitemOpen [0]{}%
\providecommand \bibitemStop [0]{}%
\providecommand \bibitemNoStop [0]{.\EOS\space}%
\providecommand \EOS [0]{\spacefactor3000\relax}%
\providecommand \BibitemShut  [1]{\csname bibitem#1\endcsname}%
\let\auto@bib@innerbib\@empty
%</preamble>
\bibitem [{\citenamefont {Will}(1993)}]{Will:1993ns}%
  \BibitemOpen
  \bibfield  {author} {\bibinfo {author} {\bibfnamefont {C.~M.}\ \bibnamefont
  {Will}},\ }\href@noop {} {\emph {\bibinfo {title} {{Theory and experiment in
  gravitational physics}}}}\ (\bibinfo {year} {1993})\BibitemShut {NoStop}%
%%CITATION = INSPIRE-357130;%%
\bibitem [{\citenamefont {Will}(2014)}]{Will_SEP}%
  \BibitemOpen
  \bibfield  {author} {\bibinfo {author} {\bibfnamefont {C.~M.}\ \bibnamefont
  {Will}},\ }\href {\doibase 10.12942/lrr-2014-4} {\bibfield  {journal}
  {\bibinfo  {journal} {Living Rev. Rel.}\ }\textbf {\bibinfo {volume} {17}},\
  \bibinfo {pages} {4} (\bibinfo {year} {2014})},\ \Eprint
  {http://arxiv.org/abs/1403.7377} {arXiv:1403.7377 [gr-qc]} \BibitemShut
  {NoStop}%
%%CITATION = ARXIV:1403.7377;%%
\bibitem [{\citenamefont {Berti}\ \emph {et~al.}(2015)\citenamefont {Berti}
  \emph {et~al.}}]{Berti_ModifiedReviewLarge}%
  \BibitemOpen
  \bibfield  {author} {\bibinfo {author} {\bibfnamefont {E.}~\bibnamefont
  {Berti}} \emph {et~al.},\ }\href {\doibase 10.1088/0264-9381/32/24/243001}
  {\bibfield  {journal} {\bibinfo  {journal} {Class. Quant. Grav.}\ }\textbf
  {\bibinfo {volume} {32}},\ \bibinfo {pages} {243001} (\bibinfo {year}
  {2015})},\ \Eprint {http://arxiv.org/abs/1501.07274} {arXiv:1501.07274
  [gr-qc]} \BibitemShut {NoStop}%
%%CITATION = ARXIV:1501.07274;%%
\bibitem [{\citenamefont {Abbott}\ \emph
  {et~al.}(2016{\natexlab{a}})\citenamefont {Abbott} \emph
  {et~al.}}]{Abbott_IMRcon2}%
  \BibitemOpen
  \bibfield  {author} {\bibinfo {author} {\bibfnamefont {B.~P.}\ \bibnamefont
  {Abbott}} \emph {et~al.} (\bibinfo {collaboration} {LIGO Scientific,
  Virgo}),\ }\href {\doibase 10.1103/PhysRevLett.116.221101,
  10.1103/PhysRevLett.121.129902} {\bibfield  {journal} {\bibinfo  {journal}
  {Phys. Rev. Lett.}\ }\textbf {\bibinfo {volume} {116}},\ \bibinfo {pages}
  {221101} (\bibinfo {year} {2016}{\natexlab{a}})},\ \bibinfo {note} {[Erratum:
  Phys. Rev. Lett.121,no.12,129902(2018)]},\ \Eprint
  {http://arxiv.org/abs/1602.03841} {arXiv:1602.03841 [gr-qc]} \BibitemShut
  {NoStop}%
%%CITATION = ARXIV:1602.03841;%%
\bibitem [{\citenamefont {Yunes}\ \emph {et~al.}(2016)\citenamefont {Yunes},
  \citenamefont {Yagi},\ and\ \citenamefont
  {Pretorius}}]{Yunes_ModifiedPhysics}%
  \BibitemOpen
  \bibfield  {author} {\bibinfo {author} {\bibfnamefont {N.}~\bibnamefont
  {Yunes}}, \bibinfo {author} {\bibfnamefont {K.}~\bibnamefont {Yagi}}, \ and\
  \bibinfo {author} {\bibfnamefont {F.}~\bibnamefont {Pretorius}},\ }\href
  {\doibase 10.1103/PhysRevD.94.084002} {\bibfield  {journal} {\bibinfo
  {journal} {Phys. Rev. D}\ }\textbf {\bibinfo {volume} {94}},\ \bibinfo
  {pages} {084002} (\bibinfo {year} {2016})}\BibitemShut {NoStop}%
\bibitem [{\citenamefont {Abbott}\ \emph
  {et~al.}(2016{\natexlab{b}})\citenamefont {Abbott} \emph
  {et~al.}}]{TheLIGOScientific:2016pea}%
  \BibitemOpen
  \bibfield  {author} {\bibinfo {author} {\bibfnamefont {B.~P.}\ \bibnamefont
  {Abbott}} \emph {et~al.} (\bibinfo {collaboration} {LIGO Scientific,
  Virgo}),\ }\href {\doibase 10.1103/PhysRevX.6.041015,
  10.1103/PhysRevX.8.039903} {\bibfield  {journal} {\bibinfo  {journal} {Phys.
  Rev.}\ }\textbf {\bibinfo {volume} {X6}},\ \bibinfo {pages} {041015}
  (\bibinfo {year} {2016}{\natexlab{b}})},\ \bibinfo {note} {[erratum: Phys.
  Rev.X8,no.3,039903(2018)]},\ \Eprint {http://arxiv.org/abs/1606.04856}
  {arXiv:1606.04856 [gr-qc]} \BibitemShut {NoStop}%
%%CITATION = ARXIV:1606.04856;%%
\bibitem [{\citenamefont {Abbott}\ \emph
  {et~al.}(2019{\natexlab{a}})\citenamefont {Abbott} \emph
  {et~al.}}]{Abbott:2018lct}%
  \BibitemOpen
  \bibfield  {author} {\bibinfo {author} {\bibfnamefont {B.~P.}\ \bibnamefont
  {Abbott}} \emph {et~al.} (\bibinfo {collaboration} {LIGO Scientific,
  Virgo}),\ }\href {\doibase 10.1103/PhysRevLett.123.011102} {\bibfield
  {journal} {\bibinfo  {journal} {Phys. Rev. Lett.}\ }\textbf {\bibinfo
  {volume} {123}},\ \bibinfo {pages} {011102} (\bibinfo {year}
  {2019}{\natexlab{a}})},\ \Eprint {http://arxiv.org/abs/1811.00364}
  {arXiv:1811.00364 [gr-qc]} \BibitemShut {NoStop}%
%%CITATION = ARXIV:1811.00364;%%
\bibitem [{\citenamefont {Abbott}\ \emph
  {et~al.}(2019{\natexlab{b}})\citenamefont {Abbott} \emph
  {et~al.}}]{LIGOScientific:2019fpa}%
  \BibitemOpen
  \bibfield  {author} {\bibinfo {author} {\bibfnamefont {B.~P.}\ \bibnamefont
  {Abbott}} \emph {et~al.} (\bibinfo {collaboration} {LIGO Scientific,
  Virgo}),\ }\href {\doibase 10.1103/PhysRevD.100.104036} {\bibfield  {journal}
  {\bibinfo  {journal} {Phys. Rev.}\ }\textbf {\bibinfo {volume} {D100}},\
  \bibinfo {pages} {104036} (\bibinfo {year} {2019}{\natexlab{b}})},\ \Eprint
  {http://arxiv.org/abs/1903.04467} {arXiv:1903.04467 [gr-qc]} \BibitemShut
  {NoStop}%
%%CITATION = ARXIV:1903.04467;%%
\bibitem [{\citenamefont {Berti}\ \emph
  {et~al.}(2018{\natexlab{a}})\citenamefont {Berti}, \citenamefont {Yagi},\
  and\ \citenamefont {Yunes}}]{Berti:2018cxi}%
  \BibitemOpen
  \bibfield  {author} {\bibinfo {author} {\bibfnamefont {E.}~\bibnamefont
  {Berti}}, \bibinfo {author} {\bibfnamefont {K.}~\bibnamefont {Yagi}}, \ and\
  \bibinfo {author} {\bibfnamefont {N.}~\bibnamefont {Yunes}},\ }\href
  {\doibase 10.1007/s10714-018-2362-8} {\bibfield  {journal} {\bibinfo
  {journal} {Gen. Rel. Grav.}\ }\textbf {\bibinfo {volume} {50}},\ \bibinfo
  {pages} {46} (\bibinfo {year} {2018}{\natexlab{a}})},\ \Eprint
  {http://arxiv.org/abs/1801.03208} {arXiv:1801.03208 [gr-qc]} \BibitemShut
  {NoStop}%
%%CITATION = ARXIV:1801.03208;%%
\bibitem [{\citenamefont {Berti}\ \emph
  {et~al.}(2018{\natexlab{b}})\citenamefont {Berti}, \citenamefont {Yagi},
  \citenamefont {Yang},\ and\ \citenamefont {Yunes}}]{Berti:2018vdi}%
  \BibitemOpen
  \bibfield  {author} {\bibinfo {author} {\bibfnamefont {E.}~\bibnamefont
  {Berti}}, \bibinfo {author} {\bibfnamefont {K.}~\bibnamefont {Yagi}},
  \bibinfo {author} {\bibfnamefont {H.}~\bibnamefont {Yang}}, \ and\ \bibinfo
  {author} {\bibfnamefont {N.}~\bibnamefont {Yunes}},\ }\href {\doibase
  10.1007/s10714-018-2372-6} {\bibfield  {journal} {\bibinfo  {journal} {Gen.
  Rel. Grav.}\ }\textbf {\bibinfo {volume} {50}},\ \bibinfo {pages} {49}
  (\bibinfo {year} {2018}{\natexlab{b}})},\ \Eprint
  {http://arxiv.org/abs/1801.03587} {arXiv:1801.03587 [gr-qc]} \BibitemShut
  {NoStop}%
%%CITATION = ARXIV:1801.03587;%%
\bibitem [{\citenamefont {Clifton}\ \emph {et~al.}(2012)\citenamefont
  {Clifton}, \citenamefont {Ferreira}, \citenamefont {Padilla},\ and\
  \citenamefont {Skordis}}]{Clifton:2011jh}%
  \BibitemOpen
  \bibfield  {author} {\bibinfo {author} {\bibfnamefont {T.}~\bibnamefont
  {Clifton}}, \bibinfo {author} {\bibfnamefont {P.~G.}\ \bibnamefont
  {Ferreira}}, \bibinfo {author} {\bibfnamefont {A.}~\bibnamefont {Padilla}}, \
  and\ \bibinfo {author} {\bibfnamefont {C.}~\bibnamefont {Skordis}},\ }\href
  {\doibase 10.1016/j.physrep.2012.01.001} {\bibfield  {journal} {\bibinfo
  {journal} {Phys.Rept.}\ }\textbf {\bibinfo {volume} {513}},\ \bibinfo {pages}
  {1} (\bibinfo {year} {2012})},\ \Eprint {http://arxiv.org/abs/1106.2476}
  {arXiv:1106.2476 [astro-ph.CO]} \BibitemShut {NoStop}%
%%CITATION = ARXIV:1106.2476;%%
\bibitem [{\citenamefont {Jain}\ and\ \citenamefont
  {Khoury}(2010)}]{Jain:2010ka}%
  \BibitemOpen
  \bibfield  {author} {\bibinfo {author} {\bibfnamefont {B.}~\bibnamefont
  {Jain}}\ and\ \bibinfo {author} {\bibfnamefont {J.}~\bibnamefont {Khoury}},\
  }\href {\doibase 10.1016/j.aop.2010.04.002} {\bibfield  {journal} {\bibinfo
  {journal} {Annals Phys.}\ }\textbf {\bibinfo {volume} {325}},\ \bibinfo
  {pages} {1479} (\bibinfo {year} {2010})},\ \Eprint
  {http://arxiv.org/abs/1004.3294} {arXiv:1004.3294 [astro-ph.CO]} \BibitemShut
  {NoStop}%
%%CITATION = ARXIV:1004.3294;%%
\bibitem [{\citenamefont {Joyce}\ \emph {et~al.}(2015)\citenamefont {Joyce},
  \citenamefont {Jain}, \citenamefont {Khoury},\ and\ \citenamefont
  {Trodden}}]{Joyce:2014kja}%
  \BibitemOpen
  \bibfield  {author} {\bibinfo {author} {\bibfnamefont {A.}~\bibnamefont
  {Joyce}}, \bibinfo {author} {\bibfnamefont {B.}~\bibnamefont {Jain}},
  \bibinfo {author} {\bibfnamefont {J.}~\bibnamefont {Khoury}}, \ and\ \bibinfo
  {author} {\bibfnamefont {M.}~\bibnamefont {Trodden}},\ }\href {\doibase
  10.1016/j.physrep.2014.12.002} {\bibfield  {journal} {\bibinfo  {journal}
  {Phys. Rept.}\ }\textbf {\bibinfo {volume} {568}},\ \bibinfo {pages} {1}
  (\bibinfo {year} {2015})},\ \Eprint {http://arxiv.org/abs/1407.0059}
  {arXiv:1407.0059 [astro-ph.CO]} \BibitemShut {NoStop}%
%%CITATION = ARXIV:1407.0059;%%
\bibitem [{\citenamefont {Koyama}(2016)}]{Koyama:2015vza}%
  \BibitemOpen
  \bibfield  {author} {\bibinfo {author} {\bibfnamefont {K.}~\bibnamefont
  {Koyama}},\ }\href {\doibase 10.1088/0034-4885/79/4/046902} {\bibfield
  {journal} {\bibinfo  {journal} {Rept. Prog. Phys.}\ }\textbf {\bibinfo
  {volume} {79}},\ \bibinfo {pages} {046902} (\bibinfo {year} {2016})},\
  \Eprint {http://arxiv.org/abs/1504.04623} {arXiv:1504.04623 [astro-ph.CO]}
  \BibitemShut {NoStop}%
%%CITATION = ARXIV:1504.04623;%%
\bibitem [{\citenamefont {Amendola}\ and\ \citenamefont
  {Tsujikawa}(2010)}]{DEBook}%
  \BibitemOpen
  \bibfield  {author} {\bibinfo {author} {\bibfnamefont {L.}~\bibnamefont
  {Amendola}}\ and\ \bibinfo {author} {\bibfnamefont {S.}~\bibnamefont
  {Tsujikawa}},\ }\href@noop {} {\emph {\bibinfo {title} {Dark energy. Theory
  and Observations}}}\ (\bibinfo  {publisher} {Cambridge University Press},\
  \bibinfo {year} {2010})\BibitemShut {NoStop}%
\bibitem [{\citenamefont {Horndeski}(1974)}]{Horndeski:1974wa}%
  \BibitemOpen
  \bibfield  {author} {\bibinfo {author} {\bibfnamefont {G.~W.}\ \bibnamefont
  {Horndeski}},\ }\href {\doibase 10.1007/BF01807638} {\bibfield  {journal}
  {\bibinfo  {journal} {Int. J. Theor. Phys.}\ }\textbf {\bibinfo {volume}
  {10}},\ \bibinfo {pages} {363} (\bibinfo {year} {1974})}\BibitemShut
  {NoStop}%
%%CITATION = IJTPB,10,363;%%
\bibitem [{\citenamefont {Deffayet}\ \emph {et~al.}(2011)\citenamefont
  {Deffayet}, \citenamefont {Gao}, \citenamefont {Steer},\ and\ \citenamefont
  {Zahariade}}]{Deffayet:2011gz}%
  \BibitemOpen
  \bibfield  {author} {\bibinfo {author} {\bibfnamefont {C.}~\bibnamefont
  {Deffayet}}, \bibinfo {author} {\bibfnamefont {X.}~\bibnamefont {Gao}},
  \bibinfo {author} {\bibfnamefont {D.~A.}\ \bibnamefont {Steer}}, \ and\
  \bibinfo {author} {\bibfnamefont {G.}~\bibnamefont {Zahariade}},\ }\href
  {\doibase 10.1103/PhysRevD.84.064039} {\bibfield  {journal} {\bibinfo
  {journal} {Phys. Rev.}\ }\textbf {\bibinfo {volume} {D84}},\ \bibinfo {pages}
  {064039} (\bibinfo {year} {2011})},\ \Eprint {http://arxiv.org/abs/1103.3260}
  {arXiv:1103.3260 [hep-th]} \BibitemShut {NoStop}%
%%CITATION = ARXIV:1103.3260;%%
\bibitem [{\citenamefont {Gleyzes}\ \emph {et~al.}(2015)\citenamefont
  {Gleyzes}, \citenamefont {Langlois}, \citenamefont {Piazza},\ and\
  \citenamefont {Vernizzi}}]{Gleyzes:2014dya}%
  \BibitemOpen
  \bibfield  {author} {\bibinfo {author} {\bibfnamefont {J.}~\bibnamefont
  {Gleyzes}}, \bibinfo {author} {\bibfnamefont {D.}~\bibnamefont {Langlois}},
  \bibinfo {author} {\bibfnamefont {F.}~\bibnamefont {Piazza}}, \ and\ \bibinfo
  {author} {\bibfnamefont {F.}~\bibnamefont {Vernizzi}},\ }\href {\doibase
  10.1103/PhysRevLett.114.211101} {\bibfield  {journal} {\bibinfo  {journal}
  {Phys. Rev. Lett.}\ }\textbf {\bibinfo {volume} {114}},\ \bibinfo {pages}
  {211101} (\bibinfo {year} {2015})},\ \Eprint {http://arxiv.org/abs/1404.6495}
  {arXiv:1404.6495 [hep-th]} \BibitemShut {NoStop}%
%%CITATION = ARXIV:1404.6495;%%
\bibitem [{\citenamefont {Kobayashi}(2019)}]{Kobayashi:2019hrl}%
  \BibitemOpen
  \bibfield  {author} {\bibinfo {author} {\bibfnamefont {T.}~\bibnamefont
  {Kobayashi}},\ }\href {\doibase 10.1088/1361-6633/ab2429} {\bibfield
  {journal} {\bibinfo  {journal} {Rept. Prog. Phys.}\ }\textbf {\bibinfo
  {volume} {82}},\ \bibinfo {pages} {086901} (\bibinfo {year} {2019})},\
  \Eprint {http://arxiv.org/abs/1901.07183} {arXiv:1901.07183 [gr-qc]}
  \BibitemShut {NoStop}%
%%CITATION = ARXIV:1901.07183;%%
\bibitem [{\citenamefont {Horava}(2009)}]{Horava:2009uw}%
  \BibitemOpen
  \bibfield  {author} {\bibinfo {author} {\bibfnamefont {P.}~\bibnamefont
  {Horava}},\ }\href {\doibase 10.1103/PhysRevD.79.084008} {\bibfield
  {journal} {\bibinfo  {journal} {Phys. Rev.}\ }\textbf {\bibinfo {volume}
  {D79}},\ \bibinfo {pages} {084008} (\bibinfo {year} {2009})},\ \Eprint
  {http://arxiv.org/abs/0901.3775} {arXiv:0901.3775 [hep-th]} \BibitemShut
  {NoStop}%
%%CITATION = ARXIV:0901.3775;%%
\bibitem [{\citenamefont {Blas}\ \emph {et~al.}(2010)\citenamefont {Blas},
  \citenamefont {Pujolas},\ and\ \citenamefont {Sibiryakov}}]{Blas:2009qj}%
  \BibitemOpen
  \bibfield  {author} {\bibinfo {author} {\bibfnamefont {D.}~\bibnamefont
  {Blas}}, \bibinfo {author} {\bibfnamefont {O.}~\bibnamefont {Pujolas}}, \
  and\ \bibinfo {author} {\bibfnamefont {S.}~\bibnamefont {Sibiryakov}},\
  }\href {\doibase 10.1103/PhysRevLett.104.181302} {\bibfield  {journal}
  {\bibinfo  {journal} {Phys. Rev. Lett.}\ }\textbf {\bibinfo {volume} {104}},\
  \bibinfo {pages} {181302} (\bibinfo {year} {2010})},\ \Eprint
  {http://arxiv.org/abs/0909.3525} {arXiv:0909.3525 [hep-th]} \BibitemShut
  {NoStop}%
%%CITATION = ARXIV:0909.3525;%%
\bibitem [{\citenamefont {Blas}\ \emph {et~al.}(2011)\citenamefont {Blas},
  \citenamefont {Pujolas},\ and\ \citenamefont {Sibiryakov}}]{Blas:2010hb}%
  \BibitemOpen
  \bibfield  {author} {\bibinfo {author} {\bibfnamefont {D.}~\bibnamefont
  {Blas}}, \bibinfo {author} {\bibfnamefont {O.}~\bibnamefont {Pujolas}}, \
  and\ \bibinfo {author} {\bibfnamefont {S.}~\bibnamefont {Sibiryakov}},\
  }\href {\doibase 10.1007/JHEP04(2011)018} {\bibfield  {journal} {\bibinfo
  {journal} {JHEP}\ }\textbf {\bibinfo {volume} {04}},\ \bibinfo {pages} {018}
  (\bibinfo {year} {2011})},\ \Eprint {http://arxiv.org/abs/1007.3503}
  {arXiv:1007.3503 [hep-th]} \BibitemShut {NoStop}%
%%CITATION = ARXIV:1007.3503;%%
\bibitem [{\citenamefont {Blas}\ and\ \citenamefont
  {Sibiryakov}(2015)}]{Blas:2014ira}%
  \BibitemOpen
  \bibfield  {author} {\bibinfo {author} {\bibfnamefont {D.}~\bibnamefont
  {Blas}}\ and\ \bibinfo {author} {\bibfnamefont {S.}~\bibnamefont
  {Sibiryakov}},\ }\href {\doibase 10.7868/S0044451015030180,
  10.1134/S1063776115030164} {\bibfield  {journal} {\bibinfo  {journal} {Zh.
  Eksp. Teor. Fiz.}\ }\textbf {\bibinfo {volume} {147}},\ \bibinfo {pages}
  {578} (\bibinfo {year} {2015})},\ \bibinfo {note} {[J. Exp. Theor.
  Phys.120,no.3,509(2015)]},\ \Eprint {http://arxiv.org/abs/1410.2408}
  {arXiv:1410.2408 [hep-th]} \BibitemShut {NoStop}%
%%CITATION = ARXIV:1410.2408;%%
\bibitem [{\citenamefont {Gabadadze}\ \emph {et~al.}(2013)\citenamefont
  {Gabadadze}, \citenamefont {Hinterbichler}, \citenamefont {Pirtskhalava},\
  and\ \citenamefont {Shang}}]{Gabadadze:2013ria}%
  \BibitemOpen
  \bibfield  {author} {\bibinfo {author} {\bibfnamefont {G.}~\bibnamefont
  {Gabadadze}}, \bibinfo {author} {\bibfnamefont {K.}~\bibnamefont
  {Hinterbichler}}, \bibinfo {author} {\bibfnamefont {D.}~\bibnamefont
  {Pirtskhalava}}, \ and\ \bibinfo {author} {\bibfnamefont {Y.}~\bibnamefont
  {Shang}},\ }\href {\doibase 10.1103/PhysRevD.88.084003} {\bibfield  {journal}
  {\bibinfo  {journal} {Phys. Rev.}\ }\textbf {\bibinfo {volume} {D88}},\
  \bibinfo {pages} {084003} (\bibinfo {year} {2013})},\ \Eprint
  {http://arxiv.org/abs/1307.2245} {arXiv:1307.2245 [hep-th]} \BibitemShut
  {NoStop}%
%%CITATION = ARXIV:1307.2245;%%
\bibitem [{\citenamefont {Ondo}\ and\ \citenamefont
  {Tolley}(2013)}]{Ondo:2013wka}%
  \BibitemOpen
  \bibfield  {author} {\bibinfo {author} {\bibfnamefont {N.~A.}\ \bibnamefont
  {Ondo}}\ and\ \bibinfo {author} {\bibfnamefont {A.~J.}\ \bibnamefont
  {Tolley}},\ }\href {\doibase 10.1007/JHEP11(2013)059} {\bibfield  {journal}
  {\bibinfo  {journal} {JHEP}\ }\textbf {\bibinfo {volume} {11}},\ \bibinfo
  {pages} {059} (\bibinfo {year} {2013})},\ \Eprint
  {http://arxiv.org/abs/1307.4769} {arXiv:1307.4769 [hep-th]} \BibitemShut
  {NoStop}%
%%CITATION = ARXIV:1307.4769;%%
\bibitem [{\citenamefont {Liberati}(2013)}]{Liberati:2013xla}%
  \BibitemOpen
  \bibfield  {author} {\bibinfo {author} {\bibfnamefont {S.}~\bibnamefont
  {Liberati}},\ }\href {\doibase 10.1088/0264-9381/30/13/133001} {\bibfield
  {journal} {\bibinfo  {journal} {Class. Quant. Grav.}\ }\textbf {\bibinfo
  {volume} {30}},\ \bibinfo {pages} {133001} (\bibinfo {year} {2013})},\
  \Eprint {http://arxiv.org/abs/1304.5795} {arXiv:1304.5795 [gr-qc]}
  \BibitemShut {NoStop}%
%%CITATION = ARXIV:1304.5795;%%
\bibitem [{\citenamefont {Yagi}\ \emph
  {et~al.}(2014{\natexlab{a}})\citenamefont {Yagi}, \citenamefont {Blas},
  \citenamefont {Yunes},\ and\ \citenamefont {Barausse}}]{Yagi:2013qpa}%
  \BibitemOpen
  \bibfield  {author} {\bibinfo {author} {\bibfnamefont {K.}~\bibnamefont
  {Yagi}}, \bibinfo {author} {\bibfnamefont {D.}~\bibnamefont {Blas}}, \bibinfo
  {author} {\bibfnamefont {N.}~\bibnamefont {Yunes}}, \ and\ \bibinfo {author}
  {\bibfnamefont {E.}~\bibnamefont {Barausse}},\ }\href {\doibase
  10.1103/PhysRevLett.112.161101} {\bibfield  {journal} {\bibinfo  {journal}
  {Phys. Rev. Lett.}\ }\textbf {\bibinfo {volume} {112}},\ \bibinfo {pages}
  {161101} (\bibinfo {year} {2014}{\natexlab{a}})},\ \Eprint
  {http://arxiv.org/abs/1307.6219} {arXiv:1307.6219 [gr-qc]} \BibitemShut
  {NoStop}%
%%CITATION = ARXIV:1307.6219;%%
\bibitem [{\citenamefont {Yagi}\ \emph
  {et~al.}(2014{\natexlab{b}})\citenamefont {Yagi}, \citenamefont {Blas},
  \citenamefont {Barausse},\ and\ \citenamefont {Yunes}}]{Yagi:2013ava}%
  \BibitemOpen
  \bibfield  {author} {\bibinfo {author} {\bibfnamefont {K.}~\bibnamefont
  {Yagi}}, \bibinfo {author} {\bibfnamefont {D.}~\bibnamefont {Blas}}, \bibinfo
  {author} {\bibfnamefont {E.}~\bibnamefont {Barausse}}, \ and\ \bibinfo
  {author} {\bibfnamefont {N.}~\bibnamefont {Yunes}},\ }\href {\doibase
  10.1103/PhysRevD.90.069902, 10.1103/PhysRevD.90.069901,
  10.1103/PhysRevD.89.084067} {\bibfield  {journal} {\bibinfo  {journal} {Phys.
  Rev.}\ }\textbf {\bibinfo {volume} {D89}},\ \bibinfo {pages} {084067}
  (\bibinfo {year} {2014}{\natexlab{b}})},\ \bibinfo {note} {[Erratum: Phys.
  Rev.D90,no.6,069901(2014)]},\ \Eprint {http://arxiv.org/abs/1311.7144}
  {arXiv:1311.7144 [gr-qc]} \BibitemShut {NoStop}%
%%CITATION = ARXIV:1311.7144;%%
\bibitem [{\citenamefont {Emir~Gümrükçüoğlu}\ \emph
  {et~al.}(2018)\citenamefont {Emir~Gümrükçüoğlu}, \citenamefont
  {Saravani},\ and\ \citenamefont {Sotiriou}}]{Gumrukcuoglu:2017ijh}%
  \BibitemOpen
  \bibfield  {author} {\bibinfo {author} {\bibfnamefont {A.}~\bibnamefont
  {Emir~Gümrükçüoğlu}}, \bibinfo {author} {\bibfnamefont {M.}~\bibnamefont
  {Saravani}}, \ and\ \bibinfo {author} {\bibfnamefont {T.~P.}\ \bibnamefont
  {Sotiriou}},\ }\href {\doibase 10.1103/PhysRevD.97.024032} {\bibfield
  {journal} {\bibinfo  {journal} {Phys. Rev.}\ }\textbf {\bibinfo {volume}
  {D97}},\ \bibinfo {pages} {024032} (\bibinfo {year} {2018})},\ \Eprint
  {http://arxiv.org/abs/1711.08845} {arXiv:1711.08845 [gr-qc]} \BibitemShut
  {NoStop}%
%%CITATION = ARXIV:1711.08845;%%
\bibitem [{\citenamefont {Oost}\ \emph {et~al.}(2018)\citenamefont {Oost},
  \citenamefont {Mukohyama},\ and\ \citenamefont {Wang}}]{Oost:2018tcv}%
  \BibitemOpen
  \bibfield  {author} {\bibinfo {author} {\bibfnamefont {J.}~\bibnamefont
  {Oost}}, \bibinfo {author} {\bibfnamefont {S.}~\bibnamefont {Mukohyama}}, \
  and\ \bibinfo {author} {\bibfnamefont {A.}~\bibnamefont {Wang}},\ }\href
  {\doibase 10.1103/PhysRevD.97.124023} {\bibfield  {journal} {\bibinfo
  {journal} {Phys. Rev.}\ }\textbf {\bibinfo {volume} {D97}},\ \bibinfo {pages}
  {124023} (\bibinfo {year} {2018})},\ \Eprint
  {http://arxiv.org/abs/1802.04303} {arXiv:1802.04303 [gr-qc]} \BibitemShut
  {NoStop}%
%%CITATION = ARXIV:1802.04303;%%
\bibitem [{\citenamefont {Heisenberg}(2014)}]{Heisenberg:2014rta}%
  \BibitemOpen
  \bibfield  {author} {\bibinfo {author} {\bibfnamefont {L.}~\bibnamefont
  {Heisenberg}},\ }\href {\doibase 10.1088/1475-7516/2014/05/015} {\bibfield
  {journal} {\bibinfo  {journal} {JCAP}\ }\textbf {\bibinfo {volume} {1405}},\
  \bibinfo {pages} {015} (\bibinfo {year} {2014})},\ \Eprint
  {http://arxiv.org/abs/1402.7026} {arXiv:1402.7026 [hep-th]} \BibitemShut
  {NoStop}%
%%CITATION = ARXIV:1402.7026;%%
\bibitem [{\citenamefont {Allys}\ \emph
  {et~al.}(2016{\natexlab{a}})\citenamefont {Allys}, \citenamefont {Peter},\
  and\ \citenamefont {Rodriguez}}]{Allys:2015sht}%
  \BibitemOpen
  \bibfield  {author} {\bibinfo {author} {\bibfnamefont {E.}~\bibnamefont
  {Allys}}, \bibinfo {author} {\bibfnamefont {P.}~\bibnamefont {Peter}}, \ and\
  \bibinfo {author} {\bibfnamefont {Y.}~\bibnamefont {Rodriguez}},\ }\href
  {\doibase 10.1088/1475-7516/2016/02/004} {\bibfield  {journal} {\bibinfo
  {journal} {JCAP}\ }\textbf {\bibinfo {volume} {1602}},\ \bibinfo {pages}
  {004} (\bibinfo {year} {2016}{\natexlab{a}})},\ \Eprint
  {http://arxiv.org/abs/1511.03101} {arXiv:1511.03101 [hep-th]} \BibitemShut
  {NoStop}%
%%CITATION = ARXIV:1511.03101;%%
\bibitem [{\citenamefont {Allys}\ \emph
  {et~al.}(2016{\natexlab{b}})\citenamefont {Allys}, \citenamefont
  {Beltran~Almeida}, \citenamefont {Peter},\ and\ \citenamefont
  {Rodriguez}}]{Allys:2016jaq}%
  \BibitemOpen
  \bibfield  {author} {\bibinfo {author} {\bibfnamefont {E.}~\bibnamefont
  {Allys}}, \bibinfo {author} {\bibfnamefont {J.~P.}\ \bibnamefont
  {Beltran~Almeida}}, \bibinfo {author} {\bibfnamefont {P.}~\bibnamefont
  {Peter}}, \ and\ \bibinfo {author} {\bibfnamefont {Y.}~\bibnamefont
  {Rodriguez}},\ }\href {\doibase 10.1088/1475-7516/2016/09/026} {\bibfield
  {journal} {\bibinfo  {journal} {JCAP}\ }\textbf {\bibinfo {volume} {1609}},\
  \bibinfo {pages} {026} (\bibinfo {year} {2016}{\natexlab{b}})},\ \Eprint
  {http://arxiv.org/abs/1605.08355} {arXiv:1605.08355 [hep-th]} \BibitemShut
  {NoStop}%
%%CITATION = ARXIV:1605.08355;%%
\bibitem [{\citenamefont {Rodriguez}\ and\ \citenamefont
  {Navarro}(2017)}]{Rodriguez:2017ckc}%
  \BibitemOpen
  \bibfield  {author} {\bibinfo {author} {\bibfnamefont {Y.}~\bibnamefont
  {Rodriguez}}\ and\ \bibinfo {author} {\bibfnamefont {A.~A.}\ \bibnamefont
  {Navarro}},\ }\bibfield  {booktitle} {\emph {\bibinfo {booktitle}
  {{Proceedings, 70\&70 Classical and Quantum Gravitation Party: Meeting with
  Two Latin American Masters on Theoretical Physics: Cartagena, Colombia,
  September 28-30, 2016}}},\ }\href {\doibase 10.1088/1742-6596/831/1/012004}
  {\bibfield  {journal} {\bibinfo  {journal} {J. Phys. Conf. Ser.}\ }\textbf
  {\bibinfo {volume} {831}},\ \bibinfo {pages} {012004} (\bibinfo {year}
  {2017})},\ \Eprint {http://arxiv.org/abs/1703.01884} {arXiv:1703.01884
  [hep-th]} \BibitemShut {NoStop}%
%%CITATION = ARXIV:1703.01884;%%
\bibitem [{\citenamefont {De~Felice}\ \emph
  {et~al.}(2016{\natexlab{a}})\citenamefont {De~Felice}, \citenamefont
  {Heisenberg}, \citenamefont {Kase}, \citenamefont {Tsujikawa}, \citenamefont
  {Zhang},\ and\ \citenamefont {Zhao}}]{De_Felice_2016_2}%
  \BibitemOpen
  \bibfield  {author} {\bibinfo {author} {\bibfnamefont {A.}~\bibnamefont
  {De~Felice}}, \bibinfo {author} {\bibfnamefont {L.}~\bibnamefont
  {Heisenberg}}, \bibinfo {author} {\bibfnamefont {R.}~\bibnamefont {Kase}},
  \bibinfo {author} {\bibfnamefont {S.}~\bibnamefont {Tsujikawa}}, \bibinfo
  {author} {\bibfnamefont {Y.-l.}\ \bibnamefont {Zhang}}, \ and\ \bibinfo
  {author} {\bibfnamefont {G.-B.}\ \bibnamefont {Zhao}},\ }\href {\doibase
  10.1103/physrevd.93.104016} {\bibfield  {journal} {\bibinfo  {journal}
  {Physical Review D}\ }\textbf {\bibinfo {volume} {93}} (\bibinfo {year}
  {2016}{\natexlab{a}}),\ 10.1103/physrevd.93.104016}\BibitemShut {NoStop}%
\bibitem [{\citenamefont {Allys}\ \emph
  {et~al.}(2016{\natexlab{c}})\citenamefont {Allys}, \citenamefont {Peter},\
  and\ \citenamefont {Rodriguez}}]{Allys:2016kbq}%
  \BibitemOpen
  \bibfield  {author} {\bibinfo {author} {\bibfnamefont {E.}~\bibnamefont
  {Allys}}, \bibinfo {author} {\bibfnamefont {P.}~\bibnamefont {Peter}}, \ and\
  \bibinfo {author} {\bibfnamefont {Y.}~\bibnamefont {Rodriguez}},\ }\href
  {\doibase 10.1103/PhysRevD.94.084041} {\bibfield  {journal} {\bibinfo
  {journal} {Phys. Rev.}\ }\textbf {\bibinfo {volume} {D94}},\ \bibinfo {pages}
  {084041} (\bibinfo {year} {2016}{\natexlab{c}})},\ \Eprint
  {http://arxiv.org/abs/1609.05870} {arXiv:1609.05870 [hep-th]} \BibitemShut
  {NoStop}%
%%CITATION = ARXIV:1609.05870;%%
\bibitem [{\citenamefont {Heisenberg}\ \emph {et~al.}(2016)\citenamefont
  {Heisenberg}, \citenamefont {Kase},\ and\ \citenamefont
  {Tsujikawa}}]{Heisenberg_2016}%
  \BibitemOpen
  \bibfield  {author} {\bibinfo {author} {\bibfnamefont {L.}~\bibnamefont
  {Heisenberg}}, \bibinfo {author} {\bibfnamefont {R.}~\bibnamefont {Kase}}, \
  and\ \bibinfo {author} {\bibfnamefont {S.}~\bibnamefont {Tsujikawa}},\ }\href
  {\doibase 10.1016/j.physletb.2016.07.052} {\bibfield  {journal} {\bibinfo
  {journal} {Physics Letters B}\ }\textbf {\bibinfo {volume} {760}},\ \bibinfo
  {pages} {617–626} (\bibinfo {year} {2016})}\BibitemShut {NoStop}%
\bibitem [{\citenamefont {Gallego~Cadavid}\ and\ \citenamefont
  {Rodriguez}(2019)}]{GallegoCadavid:2019zke}%
  \BibitemOpen
  \bibfield  {author} {\bibinfo {author} {\bibfnamefont {A.}~\bibnamefont
  {Gallego~Cadavid}}\ and\ \bibinfo {author} {\bibfnamefont {Y.}~\bibnamefont
  {Rodriguez}},\ }\href {\doibase 10.1016/j.physletb.2019.134958} {\bibfield
  {journal} {\bibinfo  {journal} {Phys. Lett.}\ }\textbf {\bibinfo {volume}
  {B798}},\ \bibinfo {pages} {134958} (\bibinfo {year} {2019})},\ \Eprint
  {http://arxiv.org/abs/1905.10664} {arXiv:1905.10664 [hep-th]} \BibitemShut
  {NoStop}%
%%CITATION = ARXIV:1905.10664;%%
\bibitem [{\citenamefont {Nakamura}\ \emph {et~al.}(2018)\citenamefont
  {Nakamura}, \citenamefont {Felice}, \citenamefont {Kase},\ and\ \citenamefont
  {Tsujikawa}}]{nakamura2018constraints}%
  \BibitemOpen
  \bibfield  {author} {\bibinfo {author} {\bibfnamefont {S.}~\bibnamefont
  {Nakamura}}, \bibinfo {author} {\bibfnamefont {A.~D.}\ \bibnamefont
  {Felice}}, \bibinfo {author} {\bibfnamefont {R.}~\bibnamefont {Kase}}, \ and\
  \bibinfo {author} {\bibfnamefont {S.}~\bibnamefont {Tsujikawa}},\ }\href@noop
  {} {\enquote {\bibinfo {title} {Constraints on massive vector dark energy
  models from integrated sachs-wolfe-galaxy cross-correlations},}\ } (\bibinfo
  {year} {2018}),\ \Eprint {http://arxiv.org/abs/1811.07541} {arXiv:1811.07541
  [astro-ph.CO]} \BibitemShut {NoStop}%
\bibitem [{\citenamefont {Felice}\ \emph {et~al.}(2016)\citenamefont {Felice},
  \citenamefont {Heisenberg}, \citenamefont {Kase}, \citenamefont {Mukohyama},
  \citenamefont {Tsujikawa},\ and\ \citenamefont {Zhang}}]{Felice_2016}%
  \BibitemOpen
  \bibfield  {author} {\bibinfo {author} {\bibfnamefont {A.~D.}\ \bibnamefont
  {Felice}}, \bibinfo {author} {\bibfnamefont {L.}~\bibnamefont {Heisenberg}},
  \bibinfo {author} {\bibfnamefont {R.}~\bibnamefont {Kase}}, \bibinfo {author}
  {\bibfnamefont {S.}~\bibnamefont {Mukohyama}}, \bibinfo {author}
  {\bibfnamefont {S.}~\bibnamefont {Tsujikawa}}, \ and\ \bibinfo {author}
  {\bibfnamefont {Y.-l.}\ \bibnamefont {Zhang}},\ }\href {\doibase
  10.1088/1475-7516/2016/06/048} {\bibfield  {journal} {\bibinfo  {journal}
  {Journal of Cosmology and Astroparticle Physics}\ }\textbf {\bibinfo {volume}
  {2016}},\ \bibinfo {pages} {048–048} (\bibinfo {year} {2016})}\BibitemShut
  {NoStop}%
\bibitem [{\citenamefont {Tasinato}(2014)}]{Tasinato2014CosmicAF}%
  \BibitemOpen
  \bibfield  {author} {\bibinfo {author} {\bibfnamefont {G.}~\bibnamefont
  {Tasinato}},\ }\href@noop {} {\bibfield  {journal} {\bibinfo  {journal}
  {Journal of High Energy Physics}\ }\textbf {\bibinfo {volume} {2014}},\
  \bibinfo {pages} {1} (\bibinfo {year} {2014})}\BibitemShut {NoStop}%
\bibitem [{\citenamefont {Rodríguez}\ and\ \citenamefont
  {Navarro}(2018)}]{Rodriguez:2017wkg}%
  \BibitemOpen
  \bibfield  {author} {\bibinfo {author} {\bibfnamefont {Y.}~\bibnamefont
  {Rodríguez}}\ and\ \bibinfo {author} {\bibfnamefont {A.~A.}\ \bibnamefont
  {Navarro}},\ }\href {\doibase 10.1016/j.dark.2018.01.003} {\bibfield
  {journal} {\bibinfo  {journal} {Phys. Dark Univ.}\ }\textbf {\bibinfo
  {volume} {19}},\ \bibinfo {pages} {129} (\bibinfo {year} {2018})},\ \Eprint
  {http://arxiv.org/abs/1711.01935} {arXiv:1711.01935 [gr-qc]} \BibitemShut
  {NoStop}%
%%CITATION = ARXIV:1711.01935;%%
\bibitem [{\citenamefont {De~Felice}\ \emph
  {et~al.}(2016{\natexlab{b}})\citenamefont {De~Felice}, \citenamefont
  {Heisenberg}, \citenamefont {Kase}, \citenamefont {Mukohyama}, \citenamefont
  {Tsujikawa},\ and\ \citenamefont {Zhang}}]{De_Felice_2016}%
  \BibitemOpen
  \bibfield  {author} {\bibinfo {author} {\bibfnamefont {A.}~\bibnamefont
  {De~Felice}}, \bibinfo {author} {\bibfnamefont {L.}~\bibnamefont
  {Heisenberg}}, \bibinfo {author} {\bibfnamefont {R.}~\bibnamefont {Kase}},
  \bibinfo {author} {\bibfnamefont {S.}~\bibnamefont {Mukohyama}}, \bibinfo
  {author} {\bibfnamefont {S.}~\bibnamefont {Tsujikawa}}, \ and\ \bibinfo
  {author} {\bibfnamefont {Y.-l.}\ \bibnamefont {Zhang}},\ }\href {\doibase
  10.1103/physrevd.94.044024} {\bibfield  {journal} {\bibinfo  {journal}
  {Physical Review D}\ }\textbf {\bibinfo {volume} {94}} (\bibinfo {year}
  {2016}{\natexlab{b}}),\ 10.1103/physrevd.94.044024}\BibitemShut {NoStop}%
\bibitem [{\citenamefont {Baker}\ \emph {et~al.}(2017)\citenamefont {Baker},
  \citenamefont {Bellini}, \citenamefont {Ferreira}, \citenamefont {Lagos},
  \citenamefont {Noller},\ and\ \citenamefont {Sawicki}}]{Baker:2017hug}%
  \BibitemOpen
  \bibfield  {author} {\bibinfo {author} {\bibfnamefont {T.}~\bibnamefont
  {Baker}}, \bibinfo {author} {\bibfnamefont {E.}~\bibnamefont {Bellini}},
  \bibinfo {author} {\bibfnamefont {P.~G.}\ \bibnamefont {Ferreira}}, \bibinfo
  {author} {\bibfnamefont {M.}~\bibnamefont {Lagos}}, \bibinfo {author}
  {\bibfnamefont {J.}~\bibnamefont {Noller}}, \ and\ \bibinfo {author}
  {\bibfnamefont {I.}~\bibnamefont {Sawicki}},\ }\href {\doibase
  10.1103/PhysRevLett.119.251301} {\bibfield  {journal} {\bibinfo  {journal}
  {Phys. Rev. Lett.}\ }\textbf {\bibinfo {volume} {119}},\ \bibinfo {pages}
  {251301} (\bibinfo {year} {2017})},\ \Eprint
  {http://arxiv.org/abs/1710.06394} {arXiv:1710.06394 [astro-ph.CO]}
  \BibitemShut {NoStop}%
%%CITATION = ARXIV:1710.06394;%%
\bibitem [{\citenamefont {Ezquiaga}\ and\ \citenamefont
  {Zumalacárregui}(2018)}]{Ezquiaga:2018btd}%
  \BibitemOpen
  \bibfield  {author} {\bibinfo {author} {\bibfnamefont {J.~M.}\ \bibnamefont
  {Ezquiaga}}\ and\ \bibinfo {author} {\bibfnamefont {M.}~\bibnamefont
  {Zumalacárregui}},\ }\href {\doibase 10.3389/fspas.2018.00044} {\bibfield
  {journal} {\bibinfo  {journal} {Front. Astron. Space Sci.}\ }\textbf
  {\bibinfo {volume} {5}},\ \bibinfo {pages} {44} (\bibinfo {year} {2018})},\
  \Eprint {http://arxiv.org/abs/1807.09241} {arXiv:1807.09241 [astro-ph.CO]}
  \BibitemShut {NoStop}%
%%CITATION = ARXIV:1807.09241;%%
\bibitem [{\citenamefont {Domènech}\ \emph {et~al.}(2018)\citenamefont
  {Domènech}, \citenamefont {Mukohyama}, \citenamefont {Namba},\ and\
  \citenamefont {Papadopoulos}}]{Domenech_2018}%
  \BibitemOpen
  \bibfield  {author} {\bibinfo {author} {\bibfnamefont {G.}~\bibnamefont
  {Domènech}}, \bibinfo {author} {\bibfnamefont {S.}~\bibnamefont
  {Mukohyama}}, \bibinfo {author} {\bibfnamefont {R.}~\bibnamefont {Namba}}, \
  and\ \bibinfo {author} {\bibfnamefont {V.}~\bibnamefont {Papadopoulos}},\
  }\href {\doibase 10.1103/physrevd.98.064037} {\bibfield  {journal} {\bibinfo
  {journal} {Physical Review D}\ }\textbf {\bibinfo {volume} {98}} (\bibinfo
  {year} {2018}),\ 10.1103/physrevd.98.064037}\BibitemShut {NoStop}%
\bibitem [{\citenamefont {Kimura}\ \emph {et~al.}(2017)\citenamefont {Kimura},
  \citenamefont {Naruko},\ and\ \citenamefont {Yoshida}}]{Kimura:2016rzw}%
  \BibitemOpen
  \bibfield  {author} {\bibinfo {author} {\bibfnamefont {R.}~\bibnamefont
  {Kimura}}, \bibinfo {author} {\bibfnamefont {A.}~\bibnamefont {Naruko}}, \
  and\ \bibinfo {author} {\bibfnamefont {D.}~\bibnamefont {Yoshida}},\ }\href
  {\doibase 10.1088/1475-7516/2017/01/002} {\bibfield  {journal} {\bibinfo
  {journal} {JCAP}\ }\textbf {\bibinfo {volume} {1701}},\ \bibinfo {pages}
  {002} (\bibinfo {year} {2017})},\ \Eprint {http://arxiv.org/abs/1608.07066}
  {arXiv:1608.07066 [gr-qc]} \BibitemShut {NoStop}%
%%CITATION = ARXIV:1608.07066;%%
\bibitem [{\citenamefont {Abbott}\ \emph
  {et~al.}(2019{\natexlab{c}})\citenamefont {Abbott} \emph
  {et~al.}}]{LIGOScientific:2018mvr}%
  \BibitemOpen
  \bibfield  {author} {\bibinfo {author} {\bibfnamefont {B.~P.}\ \bibnamefont
  {Abbott}} \emph {et~al.} (\bibinfo {collaboration} {LIGO Scientific,
  Virgo}),\ }\href {\doibase 10.1103/PhysRevX.9.031040} {\bibfield  {journal}
  {\bibinfo  {journal} {Phys. Rev.}\ }\textbf {\bibinfo {volume} {X9}},\
  \bibinfo {pages} {031040} (\bibinfo {year} {2019}{\natexlab{c}})},\ \Eprint
  {http://arxiv.org/abs/1811.12907} {arXiv:1811.12907 [astro-ph.HE]}
  \BibitemShut {NoStop}%
%%CITATION = ARXIV:1811.12907;%%
\bibitem [{\citenamefont {Akiyama}\ \emph {et~al.}(2019)\citenamefont {Akiyama}
  \emph {et~al.}}]{Akiyama:2019cqa}%
  \BibitemOpen
  \bibfield  {author} {\bibinfo {author} {\bibfnamefont {K.}~\bibnamefont
  {Akiyama}} \emph {et~al.} (\bibinfo {collaboration} {Event Horizon
  Telescope}),\ }\href {\doibase 10.3847/2041-8213/ab0ec7} {\bibfield
  {journal} {\bibinfo  {journal} {Astrophys. J. Lett.}\ }\textbf {\bibinfo
  {volume} {875}},\ \bibinfo {pages} {L1} (\bibinfo {year} {2019})},\ \Eprint
  {http://arxiv.org/abs/1906.11238} {arXiv:1906.11238 [astro-ph.GA]}
  \BibitemShut {NoStop}%
%%CITATION = ARXIV:1906.11238;%%
\bibitem [{\citenamefont {Abuter}\ \emph {et~al.}(2020)\citenamefont {Abuter}
  \emph {et~al.}}]{Abuter:2020dou}%
  \BibitemOpen
  \bibfield  {author} {\bibinfo {author} {\bibfnamefont {R.}~\bibnamefont
  {Abuter}} \emph {et~al.} (\bibinfo {collaboration} {GRAVITY}),\ }\href
  {\doibase 10.1051/0004-6361/202037813} {\bibfield  {journal} {\bibinfo
  {journal} {Astron. Astrophys.}\ }\textbf {\bibinfo {volume} {636}},\ \bibinfo
  {pages} {L5} (\bibinfo {year} {2020})},\ \Eprint
  {http://arxiv.org/abs/2004.07187} {arXiv:2004.07187 [astro-ph.GA]}
  \BibitemShut {NoStop}%
%%CITATION = ARXIV:2004.07187;%%
\bibitem [{\citenamefont {Cisterna}\ \emph {et~al.}(2016)\citenamefont
  {Cisterna}, \citenamefont {Hassaine}, \citenamefont {Oliva},\ and\
  \citenamefont {Rinaldi}}]{Cisterna_2016}%
  \BibitemOpen
  \bibfield  {author} {\bibinfo {author} {\bibfnamefont {A.}~\bibnamefont
  {Cisterna}}, \bibinfo {author} {\bibfnamefont {M.}~\bibnamefont {Hassaine}},
  \bibinfo {author} {\bibfnamefont {J.}~\bibnamefont {Oliva}}, \ and\ \bibinfo
  {author} {\bibfnamefont {M.}~\bibnamefont {Rinaldi}},\ }\href {\doibase
  10.1103/physrevd.94.104039} {\bibfield  {journal} {\bibinfo  {journal}
  {Physical Review D}\ }\textbf {\bibinfo {volume} {94}} (\bibinfo {year}
  {2016}),\ 10.1103/physrevd.94.104039}\BibitemShut {NoStop}%
\bibitem [{\citenamefont {Babichev}\ \emph {et~al.}(2017)\citenamefont
  {Babichev}, \citenamefont {Charmousis},\ and\ \citenamefont
  {Hassaine}}]{Babichev_2017}%
  \BibitemOpen
  \bibfield  {author} {\bibinfo {author} {\bibfnamefont {E.}~\bibnamefont
  {Babichev}}, \bibinfo {author} {\bibfnamefont {C.}~\bibnamefont
  {Charmousis}}, \ and\ \bibinfo {author} {\bibfnamefont {M.}~\bibnamefont
  {Hassaine}},\ }\href {\doibase 10.1007/jhep05(2017)114} {\bibfield  {journal}
  {\bibinfo  {journal} {Journal of High Energy Physics}\ }\textbf {\bibinfo
  {volume} {2017}} (\bibinfo {year} {2017}),\
  10.1007/jhep05(2017)114}\BibitemShut {NoStop}%
\bibitem [{\citenamefont {Heisenberg}\ \emph {et~al.}(2017)\citenamefont
  {Heisenberg}, \citenamefont {Kase}, \citenamefont {Minamitsuji},\ and\
  \citenamefont {Tsujikawa}}]{Heisenberg_2017}%
  \BibitemOpen
  \bibfield  {author} {\bibinfo {author} {\bibfnamefont {L.}~\bibnamefont
  {Heisenberg}}, \bibinfo {author} {\bibfnamefont {R.}~\bibnamefont {Kase}},
  \bibinfo {author} {\bibfnamefont {M.}~\bibnamefont {Minamitsuji}}, \ and\
  \bibinfo {author} {\bibfnamefont {S.}~\bibnamefont {Tsujikawa}},\ }\href
  {\doibase 10.1088/1475-7516/2017/08/024} {\bibfield  {journal} {\bibinfo
  {journal} {Journal of Cosmology and Astroparticle Physics}\ }\textbf
  {\bibinfo {volume} {2017}},\ \bibinfo {pages} {024–024} (\bibinfo {year}
  {2017})}\BibitemShut {NoStop}%
\bibitem [{\citenamefont {Kase}\ \emph {et~al.}(2018)\citenamefont {Kase},
  \citenamefont {Minamitsuji}, \citenamefont {Tsujikawa},\ and\ \citenamefont
  {Zhang}}]{Kase_2018}%
  \BibitemOpen
  \bibfield  {author} {\bibinfo {author} {\bibfnamefont {R.}~\bibnamefont
  {Kase}}, \bibinfo {author} {\bibfnamefont {M.}~\bibnamefont {Minamitsuji}},
  \bibinfo {author} {\bibfnamefont {S.}~\bibnamefont {Tsujikawa}}, \ and\
  \bibinfo {author} {\bibfnamefont {Y.-l.}\ \bibnamefont {Zhang}},\ }\href
  {\doibase 10.1088/1475-7516/2018/02/048} {\bibfield  {journal} {\bibinfo
  {journal} {Journal of Cosmology and Astroparticle Physics}\ }\textbf
  {\bibinfo {volume} {2018}},\ \bibinfo {pages} {048–048} (\bibinfo {year}
  {2018})}\BibitemShut {NoStop}%
\bibitem [{\citenamefont {Filippini}\ and\ \citenamefont
  {Tasinato}(2018)}]{Filippini:2017kov}%
  \BibitemOpen
  \bibfield  {author} {\bibinfo {author} {\bibfnamefont {F.}~\bibnamefont
  {Filippini}}\ and\ \bibinfo {author} {\bibfnamefont {G.}~\bibnamefont
  {Tasinato}},\ }\href {\doibase 10.1088/1475-7516/2018/01/033} {\bibfield
  {journal} {\bibinfo  {journal} {JCAP}\ }\textbf {\bibinfo {volume} {1801}},\
  \bibinfo {pages} {033} (\bibinfo {year} {2018})},\ \Eprint
  {http://arxiv.org/abs/1709.02147} {arXiv:1709.02147 [hep-th]} \BibitemShut
  {NoStop}%
%%CITATION = ARXIV:1709.02147;%%
\bibitem [{\citenamefont {Bekenstein}(1974)}]{Bekenstein:1974sf}%
  \BibitemOpen
  \bibfield  {author} {\bibinfo {author} {\bibfnamefont {J.~D.}\ \bibnamefont
  {Bekenstein}},\ }\href {\doibase 10.1016/0003-4916(74)90124-9} {\bibfield
  {journal} {\bibinfo  {journal} {Annals Phys.}\ }\textbf {\bibinfo {volume}
  {82}},\ \bibinfo {pages} {535} (\bibinfo {year} {1974})}\BibitemShut
  {NoStop}%
%%CITATION = APNYA,82,535;%%
\bibitem [{\citenamefont {Yazadjiev}(2002)}]{Yazadjiev:2001bx}%
  \BibitemOpen
  \bibfield  {author} {\bibinfo {author} {\bibfnamefont {S.~S.}\ \bibnamefont
  {Yazadjiev}},\ }\href {\doibase 10.1103/PhysRevD.65.084023} {\bibfield
  {journal} {\bibinfo  {journal} {Phys. Rev.}\ }\textbf {\bibinfo {volume}
  {D65}},\ \bibinfo {pages} {084023} (\bibinfo {year} {2002})},\ \Eprint
  {http://arxiv.org/abs/gr-qc/0108001} {arXiv:gr-qc/0108001 [gr-qc]}
  \BibitemShut {NoStop}%
%%CITATION = GR-QC/0108001;%%
\bibitem [{\citenamefont {Faraoni}\ \emph {et~al.}(2016)\citenamefont
  {Faraoni}, \citenamefont {Hammad},\ and\ \citenamefont
  {Belknap-Keet}}]{Faraoni:2016ozb}%
  \BibitemOpen
  \bibfield  {author} {\bibinfo {author} {\bibfnamefont {V.}~\bibnamefont
  {Faraoni}}, \bibinfo {author} {\bibfnamefont {F.}~\bibnamefont {Hammad}}, \
  and\ \bibinfo {author} {\bibfnamefont {S.~D.}\ \bibnamefont {Belknap-Keet}},\
  }\href {\doibase 10.1103/PhysRevD.94.104019} {\bibfield  {journal} {\bibinfo
  {journal} {Phys. Rev.}\ }\textbf {\bibinfo {volume} {D94}},\ \bibinfo {pages}
  {104019} (\bibinfo {year} {2016})},\ \Eprint
  {http://arxiv.org/abs/1609.02783} {arXiv:1609.02783 [gr-qc]} \BibitemShut
  {NoStop}%
%%CITATION = ARXIV:1609.02783;%%
\bibitem [{\citenamefont {Faraoni}\ \emph {et~al.}(2018)\citenamefont
  {Faraoni}, \citenamefont {Çiftci},\ and\ \citenamefont
  {Belknap-Keet}}]{Faraoni:2017ecj}%
  \BibitemOpen
  \bibfield  {author} {\bibinfo {author} {\bibfnamefont {V.}~\bibnamefont
  {Faraoni}}, \bibinfo {author} {\bibfnamefont {D.~K.}\ \bibnamefont
  {Çiftci}}, \ and\ \bibinfo {author} {\bibfnamefont {S.~D.}\ \bibnamefont
  {Belknap-Keet}},\ }\href {\doibase 10.1103/PhysRevD.97.064004} {\bibfield
  {journal} {\bibinfo  {journal} {Phys. Rev.}\ }\textbf {\bibinfo {volume}
  {D97}},\ \bibinfo {pages} {064004} (\bibinfo {year} {2018})},\ \Eprint
  {http://arxiv.org/abs/1712.02205} {arXiv:1712.02205 [gr-qc]} \BibitemShut
  {NoStop}%
%%CITATION = ARXIV:1712.02205;%%
\bibitem [{\citenamefont {Bambi}\ \emph {et~al.}(2017)\citenamefont {Bambi},
  \citenamefont {Modesto},\ and\ \citenamefont {Rachwał}}]{Bambi:2016wdn}%
  \BibitemOpen
  \bibfield  {author} {\bibinfo {author} {\bibfnamefont {C.}~\bibnamefont
  {Bambi}}, \bibinfo {author} {\bibfnamefont {L.}~\bibnamefont {Modesto}}, \
  and\ \bibinfo {author} {\bibfnamefont {L.}~\bibnamefont {Rachwał}},\ }\href
  {\doibase 10.1088/1475-7516/2017/05/003} {\bibfield  {journal} {\bibinfo
  {journal} {JCAP}\ }\textbf {\bibinfo {volume} {1705}},\ \bibinfo {pages}
  {003} (\bibinfo {year} {2017})},\ \Eprint {http://arxiv.org/abs/1611.00865}
  {arXiv:1611.00865 [gr-qc]} \BibitemShut {NoStop}%
%%CITATION = ARXIV:1611.00865;%%
\bibitem [{\citenamefont {Chauvineau}(2019)}]{Chauvineau:2018zjy}%
  \BibitemOpen
  \bibfield  {author} {\bibinfo {author} {\bibfnamefont {B.}~\bibnamefont
  {Chauvineau}},\ }\href {\doibase 10.1103/PhysRevD.100.024051} {\bibfield
  {journal} {\bibinfo  {journal} {Phys. Rev.}\ }\textbf {\bibinfo {volume}
  {D100}},\ \bibinfo {pages} {024051} (\bibinfo {year} {2019})},\ \Eprint
  {http://arxiv.org/abs/1812.04934} {arXiv:1812.04934 [gr-qc]} \BibitemShut
  {NoStop}%
%%CITATION = ARXIV:1812.04934;%%
\bibitem [{\citenamefont {Ben~Achour}\ \emph {et~al.}(2020)\citenamefont
  {Ben~Achour}, \citenamefont {Liu},\ and\ \citenamefont
  {Mukohyama}}]{BenAchour:2019fdf}%
  \BibitemOpen
  \bibfield  {author} {\bibinfo {author} {\bibfnamefont {J.}~\bibnamefont
  {Ben~Achour}}, \bibinfo {author} {\bibfnamefont {H.}~\bibnamefont {Liu}}, \
  and\ \bibinfo {author} {\bibfnamefont {S.}~\bibnamefont {Mukohyama}},\ }\href
  {\doibase 10.1088/1475-7516/2020/02/023} {\bibfield  {journal} {\bibinfo
  {journal} {JCAP}\ }\textbf {\bibinfo {volume} {2002}},\ \bibinfo {pages}
  {023} (\bibinfo {year} {2020})},\ \Eprint {http://arxiv.org/abs/1910.11017}
  {arXiv:1910.11017 [gr-qc]} \BibitemShut {NoStop}%
%%CITATION = ARXIV:1910.11017;%%
\bibitem [{\citenamefont {Newman}\ \emph {et~al.}(1965)\citenamefont {Newman},
  \citenamefont {Couch}, \citenamefont {Chinnapared}, \citenamefont {Exton},
  \citenamefont {Prakash},\ and\ \citenamefont {Torrence}}]{Newman:1965my}%
  \BibitemOpen
  \bibfield  {author} {\bibinfo {author} {\bibfnamefont {E.~T.}\ \bibnamefont
  {Newman}}, \bibinfo {author} {\bibfnamefont {R.}~\bibnamefont {Couch}},
  \bibinfo {author} {\bibfnamefont {K.}~\bibnamefont {Chinnapared}}, \bibinfo
  {author} {\bibfnamefont {A.}~\bibnamefont {Exton}}, \bibinfo {author}
  {\bibfnamefont {A.}~\bibnamefont {Prakash}}, \ and\ \bibinfo {author}
  {\bibfnamefont {R.}~\bibnamefont {Torrence}},\ }\href {\doibase
  10.1063/1.1704351} {\bibfield  {journal} {\bibinfo  {journal} {J. Math.
  Phys.}\ }\textbf {\bibinfo {volume} {6}},\ \bibinfo {pages} {918} (\bibinfo
  {year} {1965})}\BibitemShut {NoStop}%
%%CITATION = JMAPA,6,918;%%
\bibitem [{\citenamefont {Thorne}(1980)}]{RevModPhys.52.299}%
  \BibitemOpen
  \bibfield  {author} {\bibinfo {author} {\bibfnamefont {K.~S.}\ \bibnamefont
  {Thorne}},\ }\href {\doibase 10.1103/RevModPhys.52.299} {\bibfield  {journal}
  {\bibinfo  {journal} {Rev. Mod. Phys.}\ }\textbf {\bibinfo {volume} {52}},\
  \bibinfo {pages} {299} (\bibinfo {year} {1980})}\BibitemShut {NoStop}%
\bibitem [{\citenamefont {Sotiriou}\ and\ \citenamefont
  {Apostolatos}(2004)}]{Sotiriou:2004ud}%
  \BibitemOpen
  \bibfield  {author} {\bibinfo {author} {\bibfnamefont {T.~P.}\ \bibnamefont
  {Sotiriou}}\ and\ \bibinfo {author} {\bibfnamefont {T.~A.}\ \bibnamefont
  {Apostolatos}},\ }\href {\doibase 10.1088/0264-9381/21/24/003} {\bibfield
  {journal} {\bibinfo  {journal} {Class. Quant. Grav.}\ }\textbf {\bibinfo
  {volume} {21}},\ \bibinfo {pages} {5727} (\bibinfo {year} {2004})},\ \Eprint
  {http://arxiv.org/abs/gr-qc/0407064} {arXiv:gr-qc/0407064 [gr-qc]}
  \BibitemShut {NoStop}%
%%CITATION = GR-QC/0407064;%%
\bibitem [{\citenamefont
  {Chandrasekhar}(1998)}]{chandrasekhar1998mathematical}%
  \BibitemOpen
  \bibfield  {author} {\bibinfo {author} {\bibfnamefont {S.}~\bibnamefont
  {Chandrasekhar}},\ }\href {https://books.google.com/books?id=LBOVcrzFfhsC}
  {\emph {\bibinfo {title} {The Mathematical Theory of Black Holes}}},\
  International series of monographs on physics\ (\bibinfo  {publisher}
  {Clarendon Press},\ \bibinfo {year} {1998})\BibitemShut {NoStop}%
\bibitem [{\citenamefont {Znajek}(1977)}]{10.1093/mnras/179.3.457}%
  \BibitemOpen
  \bibfield  {author} {\bibinfo {author} {\bibfnamefont {R.~L.}\ \bibnamefont
  {Znajek}},\ }\href {\doibase 10.1093/mnras/179.3.457} {\bibfield  {journal}
  {\bibinfo  {journal} {Monthly Notices of the Royal Astronomical Society}\
  }\textbf {\bibinfo {volume} {179}},\ \bibinfo {pages} {457} (\bibinfo {year}
  {1977})}\BibitemShut {NoStop}%
\bibitem [{\citenamefont {De~Smet}(2005)}]{DeSmet:2004if}%
  \BibitemOpen
  \bibfield  {author} {\bibinfo {author} {\bibfnamefont {P.-J.}\ \bibnamefont
  {De~Smet}},\ }\href {\doibase 10.1007/s10714-005-0013-3} {\bibfield
  {journal} {\bibinfo  {journal} {Gen. Rel. Grav.}\ }\textbf {\bibinfo {volume}
  {37}},\ \bibinfo {pages} {237} (\bibinfo {year} {2005})},\ \Eprint
  {http://arxiv.org/abs/gr-qc/0401033} {arXiv:gr-qc/0401033 [gr-qc]}
  \BibitemShut {NoStop}%
%%CITATION = GR-QC/0401033;%%
\bibitem [{\citenamefont {Stephani}\ \emph {et~al.}(2003)\citenamefont
  {Stephani}, \citenamefont {Kramer}, \citenamefont {MacCallum}, \citenamefont
  {Hoenselaers},\ and\ \citenamefont {Herlt}}]{Stephani:2003tm}%
  \BibitemOpen
  \bibfield  {author} {\bibinfo {author} {\bibfnamefont {H.}~\bibnamefont
  {Stephani}}, \bibinfo {author} {\bibfnamefont {D.}~\bibnamefont {Kramer}},
  \bibinfo {author} {\bibfnamefont {M.~A.~H.}\ \bibnamefont {MacCallum}},
  \bibinfo {author} {\bibfnamefont {C.}~\bibnamefont {Hoenselaers}}, \ and\
  \bibinfo {author} {\bibfnamefont {E.}~\bibnamefont {Herlt}},\ }\href
  {\doibase 10.1017/CBO9780511535185} {\emph {\bibinfo {title} {{Exact
  solutions of Einstein's field equations}}}},\ Cambridge Monographs on
  Mathematical Physics\ (\bibinfo  {publisher} {Cambridge Univ. Press},\
  \bibinfo {address} {Cambridge},\ \bibinfo {year} {2003})\BibitemShut
  {NoStop}%
%%CITATION = INSPIRE-619666;%%
\bibitem [{\citenamefont {Wald}(1984)}]{Wald:1984rg}%
  \BibitemOpen
  \bibfield  {author} {\bibinfo {author} {\bibfnamefont {R.~M.}\ \bibnamefont
  {Wald}},\ }\href {\doibase 10.7208/chicago/9780226870373.001.0001} {\emph
  {\bibinfo {title} {{General Relativity}}}}\ (\bibinfo  {publisher} {Chicago
  Univ. Pr.},\ \bibinfo {address} {Chicago, USA},\ \bibinfo {year}
  {1984})\BibitemShut {NoStop}%
%%CITATION = INSPIRE-209356;%%
\bibitem [{\citenamefont {Walker}\ and\ \citenamefont
  {Penrose}(1970)}]{walker1970}%
  \BibitemOpen
  \bibfield  {author} {\bibinfo {author} {\bibfnamefont {M.}~\bibnamefont
  {Walker}}\ and\ \bibinfo {author} {\bibfnamefont {R.}~\bibnamefont
  {Penrose}},\ }\href {https://projecteuclid.org:443/euclid.cmp/1103842577}
  {\bibfield  {journal} {\bibinfo  {journal} {Comm. Math. Phys.}\ }\textbf
  {\bibinfo {volume} {18}},\ \bibinfo {pages} {265} (\bibinfo {year}
  {1970})}\BibitemShut {NoStop}%
\bibitem [{\citenamefont {Benenti}\ and\ \citenamefont
  {Francaviglia}(1979)}]{Benenti}%
  \BibitemOpen
  \bibfield  {author} {\bibinfo {author} {\bibfnamefont {S.}~\bibnamefont
  {Benenti}}\ and\ \bibinfo {author} {\bibfnamefont {M.}~\bibnamefont
  {Francaviglia}},\ }\href {\doibase https://doi.org/10.1007/BF00757025}
  {\bibfield  {journal} {\bibinfo  {journal} {Gen Relat Gravit}\ }\textbf
  {\bibinfo {volume} {10}} (\bibinfo {year} {1979}),\
  https://doi.org/10.1007/BF00757025}\BibitemShut {NoStop}%
\bibitem [{\citenamefont {Vigeland}\ \emph {et~al.}(2011)\citenamefont
  {Vigeland}, \citenamefont {Yunes},\ and\ \citenamefont
  {Stein}}]{Vigeland:2011ji}%
  \BibitemOpen
  \bibfield  {author} {\bibinfo {author} {\bibfnamefont {S.}~\bibnamefont
  {Vigeland}}, \bibinfo {author} {\bibfnamefont {N.}~\bibnamefont {Yunes}}, \
  and\ \bibinfo {author} {\bibfnamefont {L.}~\bibnamefont {Stein}},\ }\href
  {\doibase 10.1103/PhysRevD.83.104027} {\bibfield  {journal} {\bibinfo
  {journal} {Phys. Rev.}\ }\textbf {\bibinfo {volume} {D83}},\ \bibinfo {pages}
  {104027} (\bibinfo {year} {2011})},\ \Eprint {http://arxiv.org/abs/1102.3706}
  {arXiv:1102.3706 [gr-qc]} \BibitemShut {NoStop}%
%%CITATION = ARXIV:1102.3706;%%
\bibitem [{\citenamefont {Johannsen}(2013)}]{Johannsen:2015pca}%
  \BibitemOpen
  \bibfield  {author} {\bibinfo {author} {\bibfnamefont {T.}~\bibnamefont
  {Johannsen}},\ }\href {\doibase 10.1103/PhysRevD.88.044002} {\bibfield
  {journal} {\bibinfo  {journal} {Phys. Rev.}\ }\textbf {\bibinfo {volume}
  {D88}},\ \bibinfo {pages} {044002} (\bibinfo {year} {2013})},\ \Eprint
  {http://arxiv.org/abs/1501.02809} {arXiv:1501.02809 [gr-qc]} \BibitemShut
  {NoStop}%
%%CITATION = ARXIV:1501.02809;%%
\bibitem [{\citenamefont {Konoplya}\ \emph {et~al.}(2018)\citenamefont
  {Konoplya}, \citenamefont {Stuchlík},\ and\ \citenamefont
  {Zhidenko}}]{Konoplya:2018arm}%
  \BibitemOpen
  \bibfield  {author} {\bibinfo {author} {\bibfnamefont {R.~A.}\ \bibnamefont
  {Konoplya}}, \bibinfo {author} {\bibfnamefont {Z.}~\bibnamefont {Stuchlík}},
  \ and\ \bibinfo {author} {\bibfnamefont {A.}~\bibnamefont {Zhidenko}},\
  }\href {\doibase 10.1103/PhysRevD.97.084044} {\bibfield  {journal} {\bibinfo
  {journal} {Phys. Rev.}\ }\textbf {\bibinfo {volume} {D97}},\ \bibinfo {pages}
  {084044} (\bibinfo {year} {2018})},\ \Eprint
  {http://arxiv.org/abs/1801.07195} {arXiv:1801.07195 [gr-qc]} \BibitemShut
  {NoStop}%
%%CITATION = ARXIV:1801.07195;%%
\bibitem [{\citenamefont {Papadopoulos}\ and\ \citenamefont
  {Kokkotas}(2018)}]{Papadopoulos:2018nvd}%
  \BibitemOpen
  \bibfield  {author} {\bibinfo {author} {\bibfnamefont {G.~O.}\ \bibnamefont
  {Papadopoulos}}\ and\ \bibinfo {author} {\bibfnamefont {K.~D.}\ \bibnamefont
  {Kokkotas}},\ }\href {\doibase 10.1088/1361-6382/aad7f4} {\bibfield
  {journal} {\bibinfo  {journal} {Class. Quant. Grav.}\ }\textbf {\bibinfo
  {volume} {35}},\ \bibinfo {pages} {185014} (\bibinfo {year} {2018})},\
  \Eprint {http://arxiv.org/abs/1807.08594} {arXiv:1807.08594 [gr-qc]}
  \BibitemShut {NoStop}%
%%CITATION = ARXIV:1807.08594;%%
\bibitem [{\citenamefont {Carson}\ and\ \citenamefont
  {Yagi}(2020)}]{Carson:2020dez}%
  \BibitemOpen
  \bibfield  {author} {\bibinfo {author} {\bibfnamefont {Z.}~\bibnamefont
  {Carson}}\ and\ \bibinfo {author} {\bibfnamefont {K.}~\bibnamefont {Yagi}},\
  }\href {\doibase 10.1103/PhysRevD.101.084030} {\bibfield  {journal} {\bibinfo
   {journal} {Phys. Rev.}\ }\textbf {\bibinfo {volume} {D101}},\ \bibinfo
  {pages} {084030} (\bibinfo {year} {2020})},\ \Eprint
  {http://arxiv.org/abs/2002.01028} {arXiv:2002.01028 [gr-qc]} \BibitemShut
  {NoStop}%
%%CITATION = ARXIV:2002.01028;%%
\bibitem [{\citenamefont {Newman}\ and\ \citenamefont
  {Janis}(1965)}]{Newman:1965tw}%
  \BibitemOpen
  \bibfield  {author} {\bibinfo {author} {\bibfnamefont {E.~T.}\ \bibnamefont
  {Newman}}\ and\ \bibinfo {author} {\bibfnamefont {A.~I.}\ \bibnamefont
  {Janis}},\ }\href {\doibase 10.1063/1.1704350} {\bibfield  {journal}
  {\bibinfo  {journal} {J. Math. Phys.}\ }\textbf {\bibinfo {volume} {6}},\
  \bibinfo {pages} {915} (\bibinfo {year} {1965})}\BibitemShut {NoStop}%
%%CITATION = JMAPA,6,915;%%
\bibitem [{\citenamefont {Drake}\ and\ \citenamefont
  {Szekeres}(2000)}]{Drake:1998gf}%
  \BibitemOpen
  \bibfield  {author} {\bibinfo {author} {\bibfnamefont {S.~P.}\ \bibnamefont
  {Drake}}\ and\ \bibinfo {author} {\bibfnamefont {P.}~\bibnamefont
  {Szekeres}},\ }\href {\doibase 10.1023/A:1001920232180} {\bibfield  {journal}
  {\bibinfo  {journal} {Gen. Rel. Grav.}\ }\textbf {\bibinfo {volume} {32}},\
  \bibinfo {pages} {445} (\bibinfo {year} {2000})},\ \Eprint
  {http://arxiv.org/abs/gr-qc/9807001} {arXiv:gr-qc/9807001 [gr-qc]}
  \BibitemShut {NoStop}%
%%CITATION = GR-QC/9807001;%%
\bibitem [{\citenamefont {Erbin}(2017)}]{Erbin:2016lzq}%
  \BibitemOpen
  \bibfield  {author} {\bibinfo {author} {\bibfnamefont {H.}~\bibnamefont
  {Erbin}},\ }\href {\doibase 10.3390/universe3010019} {\bibfield  {journal}
  {\bibinfo  {journal} {Universe}\ }\textbf {\bibinfo {volume} {3}},\ \bibinfo
  {pages} {19} (\bibinfo {year} {2017})},\ \Eprint
  {http://arxiv.org/abs/1701.00037} {arXiv:1701.00037 [gr-qc]} \BibitemShut
  {NoStop}%
%%CITATION = ARXIV:1701.00037;%%
\bibitem [{\citenamefont {Hansen}\ and\ \citenamefont
  {Yunes}(2013)}]{Hansen:2013owa}%
  \BibitemOpen
  \bibfield  {author} {\bibinfo {author} {\bibfnamefont {D.}~\bibnamefont
  {Hansen}}\ and\ \bibinfo {author} {\bibfnamefont {N.}~\bibnamefont {Yunes}},\
  }\href {\doibase 10.1103/PhysRevD.88.104020} {\bibfield  {journal} {\bibinfo
  {journal} {Phys. Rev.}\ }\textbf {\bibinfo {volume} {D88}},\ \bibinfo {pages}
  {104020} (\bibinfo {year} {2013})},\ \Eprint {http://arxiv.org/abs/1308.6631}
  {arXiv:1308.6631 [gr-qc]} \BibitemShut {NoStop}%
%%CITATION = ARXIV:1308.6631;%%
\bibitem [{\citenamefont {Yazadjiev}(2000)}]{Yazadjiev:1999ce}%
  \BibitemOpen
  \bibfield  {author} {\bibinfo {author} {\bibfnamefont {S.}~\bibnamefont
  {Yazadjiev}},\ }\href {\doibase 10.1023/A:1002080003862} {\bibfield
  {journal} {\bibinfo  {journal} {Gen. Rel. Grav.}\ }\textbf {\bibinfo {volume}
  {32}},\ \bibinfo {pages} {2345} (\bibinfo {year} {2000})},\ \Eprint
  {http://arxiv.org/abs/gr-qc/9907092} {arXiv:gr-qc/9907092 [gr-qc]}
  \BibitemShut {NoStop}%
%%CITATION = GR-QC/9907092;%%
\bibitem [{\citenamefont {Wei}\ and\ \citenamefont {Liu}(2020)}]{Wei:2020ght}%
  \BibitemOpen
  \bibfield  {author} {\bibinfo {author} {\bibfnamefont {S.-W.}\ \bibnamefont
  {Wei}}\ and\ \bibinfo {author} {\bibfnamefont {Y.-X.}\ \bibnamefont {Liu}},\
  }\href@noop {} {\  (\bibinfo {year} {2020})},\ \Eprint
  {http://arxiv.org/abs/2003.07769} {arXiv:2003.07769 [gr-qc]} \BibitemShut
  {NoStop}%
%%CITATION = ARXIV:2003.07769;%%
\bibitem [{\citenamefont {Kumar}\ and\ \citenamefont
  {Ghosh}(2020)}]{Kumar:2020owy}%
  \BibitemOpen
  \bibfield  {author} {\bibinfo {author} {\bibfnamefont {R.}~\bibnamefont
  {Kumar}}\ and\ \bibinfo {author} {\bibfnamefont {S.~G.}\ \bibnamefont
  {Ghosh}},\ }\href@noop {} {\  (\bibinfo {year} {2020})},\ \Eprint
  {http://arxiv.org/abs/2003.08927} {arXiv:2003.08927 [gr-qc]} \BibitemShut
  {NoStop}%
%%CITATION = ARXIV:2003.08927;%%
\bibitem [{\citenamefont {Erbin}(2015)}]{Erbin:2014aya}%
  \BibitemOpen
  \bibfield  {author} {\bibinfo {author} {\bibfnamefont {H.}~\bibnamefont
  {Erbin}},\ }\href {\doibase 10.1007/s10714-015-1860-1} {\bibfield  {journal}
  {\bibinfo  {journal} {Gen. Rel. Grav.}\ }\textbf {\bibinfo {volume} {47}},\
  \bibinfo {pages} {19} (\bibinfo {year} {2015})},\ \Eprint
  {http://arxiv.org/abs/1410.2602} {arXiv:1410.2602 [gr-qc]} \BibitemShut
  {NoStop}%
%%CITATION = ARXIV:1410.2602;%%
\bibitem [{\citenamefont {Maselli}\ \emph {et~al.}(2015)\citenamefont
  {Maselli}, \citenamefont {Gualtieri}, \citenamefont {Pani}, \citenamefont
  {Stella},\ and\ \citenamefont {Ferrari}}]{Maselli:2014fca}%
  \BibitemOpen
  \bibfield  {author} {\bibinfo {author} {\bibfnamefont {A.}~\bibnamefont
  {Maselli}}, \bibinfo {author} {\bibfnamefont {L.}~\bibnamefont {Gualtieri}},
  \bibinfo {author} {\bibfnamefont {P.}~\bibnamefont {Pani}}, \bibinfo {author}
  {\bibfnamefont {L.}~\bibnamefont {Stella}}, \ and\ \bibinfo {author}
  {\bibfnamefont {V.}~\bibnamefont {Ferrari}},\ }\href {\doibase
  10.1088/0004-637X/801/2/115} {\bibfield  {journal} {\bibinfo  {journal}
  {Astrophys. J.}\ }\textbf {\bibinfo {volume} {801}},\ \bibinfo {pages} {115}
  (\bibinfo {year} {2015})},\ \Eprint {http://arxiv.org/abs/1412.3473}
  {arXiv:1412.3473 [astro-ph.HE]} \BibitemShut {NoStop}%
%%CITATION = ARXIV:1412.3473;%%
\bibitem [{\citenamefont {Glampedakis}\ \emph {et~al.}(2016)\citenamefont
  {Glampedakis}, \citenamefont {Pappas}, \citenamefont {Silva},\ and\
  \citenamefont {Berti}}]{Glampedakis:2016pes}%
  \BibitemOpen
  \bibfield  {author} {\bibinfo {author} {\bibfnamefont {K.}~\bibnamefont
  {Glampedakis}}, \bibinfo {author} {\bibfnamefont {G.}~\bibnamefont {Pappas}},
  \bibinfo {author} {\bibfnamefont {H.~O.}\ \bibnamefont {Silva}}, \ and\
  \bibinfo {author} {\bibfnamefont {E.}~\bibnamefont {Berti}},\ }\href
  {\doibase 10.1103/PhysRevD.94.044030} {\bibfield  {journal} {\bibinfo
  {journal} {Phys. Rev.}\ }\textbf {\bibinfo {volume} {D94}},\ \bibinfo {pages}
  {044030} (\bibinfo {year} {2016})},\ \Eprint
  {http://arxiv.org/abs/1606.05106} {arXiv:1606.05106 [gr-qc]} \BibitemShut
  {NoStop}%
%%CITATION = ARXIV:1606.05106;%%
\bibitem [{\citenamefont {Bozzola}\ and\ \citenamefont
  {Paschalidis}(2020)}]{bozzola2020general}%
  \BibitemOpen
  \bibfield  {author} {\bibinfo {author} {\bibfnamefont {G.}~\bibnamefont
  {Bozzola}}\ and\ \bibinfo {author} {\bibfnamefont {V.}~\bibnamefont
  {Paschalidis}},\ }\href@noop {} {\enquote {\bibinfo {title} {General
  relativistic simulations of the quasi-circular inspiral and merger of charged
  black holes: Gw150914 and fundamental physics implications},}\ } (\bibinfo
  {year} {2020}),\ \Eprint {http://arxiv.org/abs/2006.15764} {arXiv:2006.15764
  [gr-qc]} \BibitemShut {NoStop}%
\bibitem [{\citenamefont {Abramowicz}\ and\ \citenamefont
  {Kluzniak}(2001)}]{Abramowicz:2001bi}%
  \BibitemOpen
  \bibfield  {author} {\bibinfo {author} {\bibfnamefont {M.~A.}\ \bibnamefont
  {Abramowicz}}\ and\ \bibinfo {author} {\bibfnamefont {W.}~\bibnamefont
  {Kluzniak}},\ }\href {\doibase 10.1051/0004-6361:20010791} {\bibfield
  {journal} {\bibinfo  {journal} {Astron. Astrophys.}\ }\textbf {\bibinfo
  {volume} {374}},\ \bibinfo {pages} {L19} (\bibinfo {year} {2001})},\ \Eprint
  {http://arxiv.org/abs/astro-ph/0105077} {arXiv:astro-ph/0105077 [astro-ph]}
  \BibitemShut {NoStop}%
%%CITATION = ASTRO-PH/0105077;%%
\bibitem [{\citenamefont {Stella}\ and\ \citenamefont
  {Vietri}(1999)}]{Stella:1998mq}%
  \BibitemOpen
  \bibfield  {author} {\bibinfo {author} {\bibfnamefont {L.}~\bibnamefont
  {Stella}}\ and\ \bibinfo {author} {\bibfnamefont {M.}~\bibnamefont
  {Vietri}},\ }\href {\doibase 10.1103/PhysRevLett.82.17} {\bibfield  {journal}
  {\bibinfo  {journal} {Phys. Rev. Lett.}\ }\textbf {\bibinfo {volume} {82}},\
  \bibinfo {pages} {17} (\bibinfo {year} {1999})},\ \Eprint
  {http://arxiv.org/abs/astro-ph/9812124} {arXiv:astro-ph/9812124 [astro-ph]}
  \BibitemShut {NoStop}%
%%CITATION = ASTRO-PH/9812124;%%
\bibitem [{\citenamefont {Saffer}\ \emph {et~al.}(2019)\citenamefont {Saffer},
  \citenamefont {Silva},\ and\ \citenamefont {Yunes}}]{Saffer:2019hqn}%
  \BibitemOpen
  \bibfield  {author} {\bibinfo {author} {\bibfnamefont {A.}~\bibnamefont
  {Saffer}}, \bibinfo {author} {\bibfnamefont {H.~O.}\ \bibnamefont {Silva}}, \
  and\ \bibinfo {author} {\bibfnamefont {N.}~\bibnamefont {Yunes}},\ }\href
  {\doibase 10.1103/PhysRevD.100.044030} {\bibfield  {journal} {\bibinfo
  {journal} {Phys. Rev.}\ }\textbf {\bibinfo {volume} {D100}},\ \bibinfo
  {pages} {044030} (\bibinfo {year} {2019})},\ \Eprint
  {http://arxiv.org/abs/1903.07779} {arXiv:1903.07779 [gr-qc]} \BibitemShut
  {NoStop}%
%%CITATION = ARXIV:1903.07779;%%
\bibitem [{\citenamefont {Gainutdinov}(2020)}]{Gainutdinov:2020bbv}%
  \BibitemOpen
  \bibfield  {author} {\bibinfo {author} {\bibfnamefont {R.~I.}\ \bibnamefont
  {Gainutdinov}},\ }\href@noop {} {\  (\bibinfo {year} {2020})},\ \Eprint
  {http://arxiv.org/abs/2002.12598} {arXiv:2002.12598 [astro-ph.GA]}
  \BibitemShut {NoStop}%
%%CITATION = ARXIV:2002.12598;%%
\bibitem [{\citenamefont {Kokkotas}(1993)}]{Kokkotas:1993ef}%
  \BibitemOpen
  \bibfield  {author} {\bibinfo {author} {\bibfnamefont {K.~D.}\ \bibnamefont
  {Kokkotas}},\ }\href {\doibase 10.1007/BF02822861} {\bibfield  {journal}
  {\bibinfo  {journal} {Nuovo Cim.}\ }\textbf {\bibinfo {volume} {B108}},\
  \bibinfo {pages} {991} (\bibinfo {year} {1993})}\BibitemShut {NoStop}%
%%CITATION = NUCIA,B108,991;%%
\bibitem [{\citenamefont {Berti}\ and\ \citenamefont
  {Kokkotas}(2005)}]{Berti:2005eb}%
  \BibitemOpen
  \bibfield  {author} {\bibinfo {author} {\bibfnamefont {E.}~\bibnamefont
  {Berti}}\ and\ \bibinfo {author} {\bibfnamefont {K.~D.}\ \bibnamefont
  {Kokkotas}},\ }\href {\doibase 10.1103/PhysRevD.71.124008} {\bibfield
  {journal} {\bibinfo  {journal} {Phys. Rev.}\ }\textbf {\bibinfo {volume}
  {D71}},\ \bibinfo {pages} {124008} (\bibinfo {year} {2005})},\ \Eprint
  {http://arxiv.org/abs/gr-qc/0502065} {arXiv:gr-qc/0502065 [gr-qc]}
  \BibitemShut {NoStop}%
%%CITATION = GR-QC/0502065;%%
\bibitem [{\citenamefont {Pani}\ \emph
  {et~al.}(2013{\natexlab{a}})\citenamefont {Pani}, \citenamefont {Berti},\
  and\ \citenamefont {Gualtieri}}]{Pani:2013ija}%
  \BibitemOpen
  \bibfield  {author} {\bibinfo {author} {\bibfnamefont {P.}~\bibnamefont
  {Pani}}, \bibinfo {author} {\bibfnamefont {E.}~\bibnamefont {Berti}}, \ and\
  \bibinfo {author} {\bibfnamefont {L.}~\bibnamefont {Gualtieri}},\ }\href
  {\doibase 10.1103/PhysRevLett.110.241103} {\bibfield  {journal} {\bibinfo
  {journal} {Phys. Rev. Lett.}\ }\textbf {\bibinfo {volume} {110}},\ \bibinfo
  {pages} {241103} (\bibinfo {year} {2013}{\natexlab{a}})},\ \Eprint
  {http://arxiv.org/abs/1304.1160} {arXiv:1304.1160 [gr-qc]} \BibitemShut
  {NoStop}%
%%CITATION = ARXIV:1304.1160;%%
\bibitem [{\citenamefont {Pani}\ \emph
  {et~al.}(2013{\natexlab{b}})\citenamefont {Pani}, \citenamefont {Berti},\
  and\ \citenamefont {Gualtieri}}]{Pani:2013wsa}%
  \BibitemOpen
  \bibfield  {author} {\bibinfo {author} {\bibfnamefont {P.}~\bibnamefont
  {Pani}}, \bibinfo {author} {\bibfnamefont {E.}~\bibnamefont {Berti}}, \ and\
  \bibinfo {author} {\bibfnamefont {L.}~\bibnamefont {Gualtieri}},\ }\href
  {\doibase 10.1103/PhysRevD.88.064048} {\bibfield  {journal} {\bibinfo
  {journal} {Phys. Rev.}\ }\textbf {\bibinfo {volume} {D88}},\ \bibinfo {pages}
  {064048} (\bibinfo {year} {2013}{\natexlab{b}})},\ \Eprint
  {http://arxiv.org/abs/1307.7315} {arXiv:1307.7315 [gr-qc]} \BibitemShut
  {NoStop}%
%%CITATION = ARXIV:1307.7315;%%
\bibitem [{\citenamefont {Mark}\ \emph {et~al.}(2015)\citenamefont {Mark},
  \citenamefont {Yang}, \citenamefont {Zimmerman},\ and\ \citenamefont
  {Chen}}]{Mark:2014aja}%
  \BibitemOpen
  \bibfield  {author} {\bibinfo {author} {\bibfnamefont {Z.}~\bibnamefont
  {Mark}}, \bibinfo {author} {\bibfnamefont {H.}~\bibnamefont {Yang}}, \bibinfo
  {author} {\bibfnamefont {A.}~\bibnamefont {Zimmerman}}, \ and\ \bibinfo
  {author} {\bibfnamefont {Y.}~\bibnamefont {Chen}},\ }\href {\doibase
  10.1103/PhysRevD.91.044025} {\bibfield  {journal} {\bibinfo  {journal} {Phys.
  Rev.}\ }\textbf {\bibinfo {volume} {D91}},\ \bibinfo {pages} {044025}
  (\bibinfo {year} {2015})},\ \Eprint {http://arxiv.org/abs/1409.5800}
  {arXiv:1409.5800 [gr-qc]} \BibitemShut {NoStop}%
%%CITATION = ARXIV:1409.5800;%%
\bibitem [{\citenamefont {Dias}\ \emph {et~al.}(2015)\citenamefont {Dias},
  \citenamefont {Godazgar},\ and\ \citenamefont {Santos}}]{Dias:2015wqa}%
  \BibitemOpen
  \bibfield  {author} {\bibinfo {author} {\bibfnamefont {O.~J.~C.}\
  \bibnamefont {Dias}}, \bibinfo {author} {\bibfnamefont {M.}~\bibnamefont
  {Godazgar}}, \ and\ \bibinfo {author} {\bibfnamefont {J.~E.}\ \bibnamefont
  {Santos}},\ }\href {\doibase 10.1103/PhysRevLett.114.151101} {\bibfield
  {journal} {\bibinfo  {journal} {Phys. Rev. Lett.}\ }\textbf {\bibinfo
  {volume} {114}},\ \bibinfo {pages} {151101} (\bibinfo {year} {2015})},\
  \Eprint {http://arxiv.org/abs/1501.04625} {arXiv:1501.04625 [gr-qc]}
  \BibitemShut {NoStop}%
%%CITATION = ARXIV:1501.04625;%%
\bibitem [{\citenamefont {Giorgi}(2020)}]{Giorgi:2020ujd}%
  \BibitemOpen
  \bibfield  {author} {\bibinfo {author} {\bibfnamefont {E.}~\bibnamefont
  {Giorgi}},\ }\href@noop {} {\  (\bibinfo {year} {2020})},\ \Eprint
  {http://arxiv.org/abs/2002.07228} {arXiv:2002.07228 [math.AP]} \BibitemShut
  {NoStop}%
%%CITATION = ARXIV:2002.07228;%%
\bibitem [{\citenamefont {Yang}\ \emph {et~al.}(2018)\citenamefont {Yang},
  \citenamefont {Ayzenberg},\ and\ \citenamefont {Bambi}}]{Yang:2018wye}%
  \BibitemOpen
  \bibfield  {author} {\bibinfo {author} {\bibfnamefont {J.}~\bibnamefont
  {Yang}}, \bibinfo {author} {\bibfnamefont {D.}~\bibnamefont {Ayzenberg}}, \
  and\ \bibinfo {author} {\bibfnamefont {C.}~\bibnamefont {Bambi}},\ }\href
  {\doibase 10.1103/PhysRevD.98.044024} {\bibfield  {journal} {\bibinfo
  {journal} {Phys. Rev.}\ }\textbf {\bibinfo {volume} {D98}},\ \bibinfo {pages}
  {044024} (\bibinfo {year} {2018})},\ \Eprint
  {http://arxiv.org/abs/1806.06240} {arXiv:1806.06240 [gr-qc]} \BibitemShut
  {NoStop}%
%%CITATION = ARXIV:1806.06240;%%
\end{thebibliography}%

\end{document}